\documentclass[prd,twocolumn,floatfix,nofootinbib]{revtex4}
\usepackage{amssymb}
\usepackage{amsmath}
\usepackage{graphicx,subfigure,color,dcolumn,booktabs,bm}
\usepackage{longtable,lscape}
\usepackage{txfonts}
\usepackage{overpic}
\usepackage{indentfirst}
\usepackage{cases}
\usepackage{multirow}
\usepackage{ulem}
\usepackage[colorlinks,
            citecolor=blue,
            anchorcolor=red,
            menucolor=red,
            linkcolor=red,
            filecolor=red,
            runcolor=red,
            urlcolor=blue,
            frenchlinks=false]{hyperref}

\graphicspath{{Figures/}}
\allowdisplaybreaks
\begin{document}

\title{Searching for compact pentaquark state within the bag model framework}
\author{Wen-Nian Liu$^{1,2}$}
\email{1169994277@qq.com}
\author{Wen-Xuan Zhang$^{1,3}$}
\author{Fu-Quan Dou$^{1,3}$\thanks{
		Corresponding author}}
\email{doufq@nwnu.edu.cn}
\affiliation{$^1$Institute of Theoretical Physics, College of Physics and Electronic
Engineering, Northwest Normal University, Lanzhou 730070, China \\
$^2$Xinjiang Laboratory Phase Transitions and Microstructures in Condensed Matters,College of Physical Science and Technology, Yili Normal University, Yining, Xinjiang,835000,China\\
$^3$Gansu Provincial Research Center for Basic Disciplines of Quantum Physics \\}
\date{\today}

\begin{abstract}
The search for the compact limit of multi-quark states is a challenging issue. Within the framework of the MIT bag model, we propose an effective limit bag radius of $R_{c} = 5.615 \, \text{GeV}^{-1}$(or $1.11 \, \text{fm}$) for bound states . When the bag radius of a hadron falls below this value, the bag binding energy satisfies $E_{B} < 0$, indicating that the system has a compact intention. We consider various combinations of different numbers and ratios of heavy and light quarks, indicating that the bag radius of hadrons depending on the number of quarks is suppressed by the presence of heavy quarks. Focusing on five-quark combinations, we find that the average bag radius of $nncc\bar{c}$ is below the threshold $R_{c}$. We take into account color-magnetic interactions and calculate the mass, magnetic moment, binding energy, and relative strong decay width for the $nnQQ\bar{Q}$ system. We show that the binding energies of most states in the flavor combination $nncc\bar{c}$ are approximately $-20 \, \text{MeV}$, whereas states involving bottom quarks have binding energies around $-120 \, \text{MeV}$, with some decay widths suppressed by the decay constant. Additionally, the $nQQQ\bar{Q}$ system exhibits even deeper binding. Our results support the compact intention of $nnQQ\bar{Q}$ and suggest that the $nQQQ\bar{Q}$ configuration demonstrates even stronger compactness.

\end{abstract}

\maketitle
\date{\today}

\section{Introduction}

At its inception, the quark model proposed the concept of exotic hadronic states that could extend beyond traditional mesons and baryons, such as tetraquarks and pentaquarks \cite{Gell-Mann:1964ewy,Zweig:1964ruk}. In the 1970s, Jaffe and others introduced the MIT Bag Model \cite{Jaffe:1976ig,Jaffe:1976ih,DeGrand:1975cf}, which has since been used to calculate double-heavy baryons \cite{He:2004px,Fleck:1989mb},  hybrid mesons \cite{Barnes:1982tx,Chanowitz:1982qj}, and light flavor pentaquark states \cite{Strottman:1979qu}.
 
 In 2003, the Belle Collaboration first reported the hidden charm tetraquark candidate $X(3872)$ \cite{Belle:2003nnu}. In the following decades, more tetraquark states were discovered sequentially, including the first charged tetraquark state $Z_c(3900)$ \cite{BESIII:2013ris} and the first tetraquark state with open charm $T_{cc}(3875)$ \cite{LHCb:2021auc}. Meanwhile, theoretical discussions have been conducted on the structure of certain tetraquarks, focusing on both molecular and compact scenarios \cite{Chen:2022asf}. Recently, the LHCb and CMS collaborations announced two fully charmed tetraquark systems, $X(6900)$ \cite{LHCb:2020bwg} and $X(6600)$ \cite{CMS:2023owd}. Many studies tend to describe fully heavy tetraquarks in compact scenarios \cite{Lu:2020cns,Tiwari:2021tmz,Liu:2021rtn,Karliner:2020dta,Faustov:2020qfm}. During this period, several signals of pentaquark states were disclosed by LHCb in the $J/\psi p$ and $\bar{p}$ channels, specifically including $P_c(4380)^+$ \cite{PhysRevLett.115.072001}, $P_c(4457)^+$, $P_c(4440)^+$, $P_c(4312)^+$\cite{LHCb:2019kea}, and $P_c(4337)$
 \cite{LHCb:2021chn}. Most studies supported the interpretation of these pentaquark states as molecular states \cite{Chen:2015sxa,Cheng:2015cca,Santopinto:2016pkp,Liu:2020hcv,Shen:2020gpw}. Subsequently, LHCb unveiled the first candidate for a strange pentaquark state, $P_{cs}(4459)^0$
 \cite{LHCb:2020jpq}, in $J/\psi\Lambda$ channel. Two years later, another strange pentaquark state, $P_{\psi s}^{\Lambda}(4338)^0$, was discovered in the same channel \cite{LHCb:2022ogu}. Research on the structures of these two strange pentaquark states mainly focuses on molecular states and compact pentaquarks \cite{Chen:2022asf,Liu:2019zoy,Chen:2016qju,Karliner:2021xnq,Wang:2019nvm,Zou:2021sha,Deng:2022vkv,Liu:2024uxn}, while also predicting other flavor combinations arising from both configurations \cite{An:2019idk,Zhang:2023hmg,Cheng:2022vgy,PhysRevD.105.034006,G:2024zkc,Wang:2019aoc}.

The interactions between molecular states and compact states differ primarily due to the varying scales of interaction involved. There are several theories suitable for compact environments, such as lattice quantum chromodynamics (lattice QCD) \cite{Meinel:2022lzo}, QCD sum rules \cite{Agaev:2021vur,Bicudo:2012qt}, color flux tube models \cite{Deng:2022vkv}, chromomagnetic interaction (CMI) models \cite{Weng:2024qly,Weng:2021ngd}, and the MIT bag model \cite{Zhang:2023teh,Jaffe:1976ih}. However, none of these theories have been used to predict the existence of ultra-large color-singlet compact hadrons. Whether multi-quark states have a compact limit is still an open question. If they do, what is the scale of this limit? Therefore, finding the compact limit of multi-quark states is an important and challenging issue.

In the present work we provide criteria for identifying compact multiquark states within the MIT bag model framework. This model effectively describes compact states; however, there is a binding upper limit for multiquark hadrons. When the bag radius of the hadron exceeds $5.615 \, \text{GeV}^{-1}$, the hadron bag will break down. Furthermore, the binding energy of the bag can be defined based on the limit radius, which allows us to evaluate the compact stability of hadrons. Taking pentaquark states as an example, we first estimate the spin-independent bag radii for the $qqqQ\bar{Q}$ and $qqQQ\bar{Q}$ states (where $q = n, s$ and $Q = c, b$). Our results indicate that for $qqqc\bar{c}$ the bag radius is slightly above the threshold, while for $qqcc\bar{c}$ it is slightly below, mainly due to the suppressive effect of heavy quarks on the bag radius. Furthermore, we calculated the mass spectra, bag radii, and compact stability for the $qqQQ\bar{Q}$ and $qQQQ\bar{Q}$ systems, with results supporting the potential for compact configurations in most states of these systems. Additionally, we discuss the two-body strong decays of the $qqQQ\bar{Q}$ system and provide branching ratio information for decay channels such as $J/\psi \Lambda_{Q}$
and $J/\psi \Xi_{Q}$.

The rest of the paper is organized as follows. In Section \ref{sec:bagmodel}, we introduce the mass formula, magnetic moment, and related parameters provided by the MIT bag model. In Section \ref{sec:compact limit}, we analyze the model and discuss the compact limit of multi-quark states. Section \ref{sec:hadrons} primarily showcases the mass spectra and magnetic moments for $qqQQ\bar{Q}$ and $qQQQ\bar{Q}$ based on the bag model, as well as the calculation results and discussion of the partial decay width ratios for the $qqQQ\bar{Q}$ system. Finally, in Section \ref{sec:summary}, we provide a concise summary of the computation results and analyses.

\section{Method for MIT bag model}
\label{sec:bagmodel}

The bag model confines the valence quarks within a sphere region characterized by radius $R$, and subsequently employs variational methods to obtain the optimal radius and mass of the hadron. Based on the traditional bag model \cite{Jaffe:1976ig,Jaffe:1976ih,DeGrand:1975cf}, Refs. \cite{Zhang:2021yul,Yan:2023lvm} presents an expression for hadron mass that incorporates the chromomagnetic interactions (CMI) between quarks as well as the non-perturbative effects. The expression is 

\begin{equation}
	M\left( R\right) =\sum_{i}\omega _{i}+\frac{4}{3}\pi R^{3}B-\frac{Z_{0}}{R}%
	+M_{B}+M_{CMI},\label{M}
\end{equation}%
where 
\begin{equation}
	\omega _{i}=\left( m_{i}^{2}+\frac{x_{i}^{2}}{R^{2}}\right) ^{1/2}.
	\label{freq}
\end{equation}%
Here, the first term on the right-hand side represents the relativistic energy sum for quarks with mass $m_{i}$, while the second and third terms correspond to the bag's volume energy and vacuum zero-point energy, with $B$ and $Z_{0}$ being model parameters; the last two terms $M_{B}$ and $M_{CMI}$ account for the short-range interactions between heavy quarks and the chromomagnetic interactions (CMI) between quarks \cite{DeRujula:1975qlm}. In Eq. (\ref{freq}), the quark momentum is given by $x_{i}/{R}$, where $x_{i}$ is a dimensionless parameter derived from the boundary conditions of the bag, satisfying the relation
\begin{equation}
	\tan x_{i}=\frac{x_{i}}{1-m_{i}R-\left( m_{i}^{2}R^{2}+x_{i}^{2}\right)
		^{1/2}},  \label{transc}
\end{equation}
which can be obtained from the spinor wave function \cite{Jaffe:1976ig,Jaffe:1976ih}. 

The short-range interactions arise from the color electric interactions resulting from the non-relativistic characteristics of heavy quarks \cite{DeGrand:1975cf,Johnson:1975zp,Karliner:2014gca,Karliner:2017elp,Karliner:2017qjm,Karliner:2020vsi}. Consequently, $M_{B}$
is related to color factors, and this contribution is mapped to the bag model in Ref. \cite{Yan:2023lvm}. For a pentaquark baryon structure with color $\bar{3}$
, the binding energy is represented by six parameters, where $B_{cc/bb}$
denotes the binding energy between doubly charm or bottom quarks, and $B_{cs/bs}$
represents the specific binding energy values, which are as follows:
\begin{equation}
	\begin{Bmatrix}
		B_{cs}=-0.025\,\text{GeV,} & B_{cc}=-0.077\,\text{GeV,} \\
		B_{bs}=-0.032\,\text{GeV,} & B_{bb}=-0.128\,\text{GeV,} \\
		B_{bc}=-0.101\,\text{GeV.} &
	\end{Bmatrix}
	\label{Bcs}
\end{equation} 

The Hamiltonian for the CMI is typically expressed by the following equation:

\begin{equation} H_{CMI} = -\sum_{i<j} \left( \mathbf{\lambda}{\mathbf{i}} \cdot \mathbf{\lambda}{j} \right) \left( \mathbf{\sigma}{\mathbf{i}} \cdot \mathbf{\sigma}{j} \right) C_{ij}, \label{CMI} 
\end{equation}
where $\mathbf{\lambda}_{\mathbf{i}}$
and $\mathbf{\sigma}_{\mathbf{i}}$
are the Gell-Mann and Pauli matrices for quark $i$, respectively, and $C_{ij}$
is the coupling constant given by \cite{DeGrand:1975cf,Yan:2023lvm}:

\begin{equation}
	\begin{aligned}
		C_{ij}=3\frac{\alpha_{s}(R)}{R^{3}}\bar{\mu}_{i}\bar{\mu}_{j}I_{ij}.
	\end{aligned}
	\label{eq:eq9}    
\end{equation}

In the bag model framework \cite{Jaffe:1976ig,Jaffe:1976ih}, the magnetic moment of a quark is
\begin{equation}
	\bar{\mu}_{i}=\frac{R}{6}\frac{4{\omega _{i}R}+2m _{i}R-3}{2{\omega
			_{i}R}\left( {\omega _{i}R}-1\right) +m _{i}R},  \label{muBari}
\end{equation}%. 
and the hadron can be expressed as 
\begin{equation}
	\mu =\left\langle \psi \left\vert \sum\nolimits_{i}2\mu
	_{i}S_{iz}\right\vert \psi \right\rangle ,  \label{musum}
\end{equation}%
where $\psi$ is the basic eigenvector of the color-spin.
 
The relevant parameters of the bag model are given in the Refs. \cite{DeGrand:1975cf,Zhang:2021yul,Zhang:2023hmg,Yan:2023lvm} as  
\begin{equation}\label{originalparas}
	\begin{Bmatrix}
		m_{n}=0\,\text{GeV,} & m_{s}=0.279\,\text{GeV,} \\
		m_{c}=1.641\,\text{GeV,} & m_{b}=5.093\,\text{GeV,} \\
		Z_{0}=1.84, & B^{1/4}=0.145\,\text{GeV.}%
	\end{Bmatrix}
\end{equation}%
The solution to the mass formula equation (\ref{M}) depends on the parameters $R$ and $x_{i}$
, where $x_{i}$ is the solution to the transcendental equation in (\ref{transc}) and is contingent upon the variational parameter $R$. Here, the values of $x_{i}$
and the bag radius $R$ can be obtained through iterative calculations.

\section{The compact limit of multi-quark states}
\label{sec:compact limit}

Next, we will discuss the binding limits described by the bag model. According to Eq. (\ref{M}), the second and third terms represent the vacuum states inside and outside the hadron, respectively. In the original Ref. \cite{DeGrand:1975cf}, the expression for the zero-point energy $E_{0} = -\frac{Z_{0}}{R}$ was derived by considering the truncation of gluon and quark fields at a finite scale, while the bag constant $B$ represents vacuum condensation in the background field. The original work also provided results for the masses, magnetic moments, and charge radii of hadrons such as $p$, $\Lambda$, $\Sigma$, $\Xi$, $\Delta$, and $\Omega$, using parameters $B^{1/4}=0.145\,\text{GeV}$ and $Z_{0}=1.84$, which matched experimental values well. Subsequent exploratory studies on heavy flavor baryons confirmed that the original bag parameters $B$ and $Z_{0}$ are reliable \cite{Zhang:2021yul,Yan:2023lvm,Zhu:2023lbx,Zhang:2023hmg}. Additionally, color confinement requires that the perturbative and non-perturbative vacuum phases does not allow for the exchange of color, which means that the hadron behaves like a cave with the boundary in a non-perturbative vacuum. To maintain this condition, we must ensure that the binding energy resulting from the vacuum effect is negative. Here, we define the bag binding energy $E_B$, 
\begin{equation}\label{EB}
E_{B}= \frac{4}{3}\pi R^{3}B - \frac{Z_{0}}{R},\\			
\end{equation}
where $Z_{0}$ and $B$ are given by Eq. (\ref{originalparas}).
When $R < 5.615 \, \text{GeV}^{-1}$
(or $1.11 \, \text{fm}$), $E_{B} < 0$, and the corresponding radius is considered as the critical value for bag confinement. 

\begin{figure}[t]
	\centering
	\includegraphics[width=0.25\textwidth]{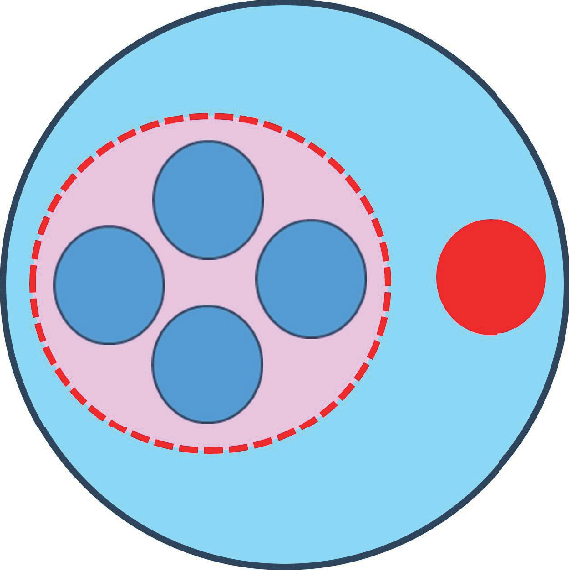}
	\caption{Multiquark states in color space. The solid red circles represent individual color triplet quarks, while the dashed circles indicate quark groups with color anti-triplet and the blue background represents the confinement region.}
	\label{fig:color}
\end{figure}

Similarly, the lattice QCD and the Schwinger model suggest the interaction limit for two quarks to be $1.13 - 1.29 \, \text{fm}$ \cite{Bulava:2019iut,Bali:2005fu,Kou:2024dml}, which is close to the result we present here. 
For compact multiquark states, we can view them as a two-body problem in color space, consisting of an arbitrarily color $3_c$ quark and a quark group with anti-color $\bar{3}_c$ (see Fig. \ref{fig:color}). Therefore, in compact states, no quark is allowed to exceed the limits of color interaction. Based on this perspective, we can extrapolate the two-body problem involving color interactions to a multi-body problem. In the bag model, we find that an increasing number of quarks leads to a continuous increase in the radius of the hadron, approaching or potentially even exceeding the bag constraint limit. Therefore, we can exclude certain multi-quark states that involve non-color interactions based on the bag binding energy or the limit radius defined by Eq. (\ref{EB}).

\renewcommand{\tabcolsep}{0.3cm} \renewcommand{\arraystretch}{0.9}
\begin{table*}[!htb]
	\caption{Ignoring the chromomagnetic interaction, we present the average bag radius and binding energy, where $n$ represents either $u$ or $d$ quarks. The unit of the bag radius $R$ is $\text{GeV}^{-1}$, while the unit of the bag binding energy is $\text{MeV}$.}
	\label{tab:radii}
	\begin{tabular}{cccccccccccc}
		\bottomrule[1.5pt]\bottomrule[0.5pt]
		System &$R$ &$E_{B}$&System &$R$ &$E_{B}$&System &$R$ &$E_{B}$&System&$R$&$E_{B}$\\ \hline
		$\bar{c}nnnn$  &$6.057$ &$108$&$n\bar{n}n\bar{n}$&$5.984$ &$89$&$nnn$&$5.223$&$-88$&$c\bar{n}$&$3.975$&$-347$\\
		$\bar{c}cnnn$  &$5.832$ &$52$&$c\bar{n}n\bar{n}$&$5.574$ &$-10$&$cnn$&$4.862$&$-166$&$c\bar{c}$&$3.442$&$-459$\\
		$\bar{c}ccnn$  &$5.491$ &$-29$&$c\bar{c}n\bar{n}$  &$5.012$ &$-134$&$ccn$&$4.422$ &$-256$&$b\bar{n}$&$3.388$&$-471$\\
		$\bar{c}cccn$  &$5.273$ &$-77$&$c\bar{c}c\bar{n}$  &$4.864$ &$-165$  &$ccc$&$4.125$ &$-316$&$b\bar{b}$&$1.948$&$-931$\\
		$\bar{b}nnnn$  &$5.924$ &$74$&$c\bar{c}c\bar{c}$&$4.608$ & $-218$&$bnn$&$4.605$&$-219$&&&\\
		$\bar{b}bnnn$  &$5.520$ &$-22$&$b\bar{n}n\bar{n}$&$5.482$ &$-31$&$bbn$&$3.854$&$-371$&&&\\
		$\bar{b}bbnn$  &$5.050$ &$-126$&$b\bar{b}n\bar{n}$&$4.701$&$-199$&$bbb$&$2.346$&$-760$&&&\\
		$\bar{b}bbbn$  &$4.308$ &$-279$&$b\bar{b}b\bar{n}$&$4.145$&$-312$&  &  &&&&\\
		$\bar{b}bbbn$  &$3.484$ &$-450$&$b\bar{b}b\bar{b}$&$3.073$&$-545$&  &  &&&&\\
		\bottomrule[0.5pt]\bottomrule[1.5pt]
	\end{tabular}%
\end{table*}

In Eq. (\ref{M}), the chromomagnetic interaction term can be regarded as a perturbation. We neglect the chromomagnetic interaction between quarks in order to calculate the bag radii of mesons, baryons, as well as tetraquark and pentaquark states, while taking into account the bag binding energy $E_B$
, as defined by Eq. (\ref{EB}). The results are presented in Table \ref{tab:radii}. The study finds that the hadron radius increases with the number of quarks, while the presence of heavy quarks has a suppressive effect on the hadron radius. This is clearly due to the fact that the incorporation of light quarks enhances the non-perturbative effects of hadrons more significantly than the influence of heavy quarks. The numerical results show that the radii of conventional baryon and meson systems are both less than the critical bag radius. For the tetraquark states, the bag radius of the all-light flavor combination $n\bar{n}n\bar{n}$ (where $n=u,d$) exceeds the binding limit, while the single heavy tetraquark $c\bar{n}n\bar{n}$ is close to the binding limit, and the doubly heavy tetraquark $c\bar{c}n\bar{n}$ has an average bag radius below the binding limit. Therefore, this does not support the existence of compact tetraquark states with all-light flavors. In the pentaquark systems, the calculated bag radii for the flavor combinations $\bar{c}cnnn$ and $\bar{c}ccnn$ are approximately $5.832 \, \text{GeV}^{-1}$ and $5.491 \, \text{GeV}^{-1}$, respectively, both of which are close to the limits for compact states. This also motivates us to further investigate triply-heavy and quadruple-heavy pentaquark states within the bag framework. 

Here, we need to clarify that in our discussion binding criticality refers to the potential for forming compact structures, while whether these structures are compact or non-compact may also relate to the hadronization processes.

 In order to check whether there is any confusion between the limit radius of compact states and the radii of molecular states, we can provide a rough estimate of the minimum radius for some molecular states using the MIT bag model. For example, the minimum radius of the $DD$ molecular state can be considered as the sum of the radii of two $D$ mesons, which is approximately $7.95 \, \text{GeV}^{-1}$ (1.57 fm). Similarly, we estimate the minimum radius of the $BB$ molecular state to be about $6.676 \, \text{GeV}^{-1}$ (1.317 fm), referring to the spin-independent bag radii of $c\bar{n}$ and $b\bar{n}$ provided in Table \ref{tab:radii}. Here, we consider the $BB$ molecular state to be the molecular state with the smallest radius, which is $6.676 \, \text{GeV}^{-1}$. Therefore, the limit radius of the compact states does not encompass the scale range of molecular states, and there exists at least a discrepancy interval of approximately $5.615 - 6.676 \, \text{GeV}^{-1}$
 .

\section{masses and decays of multi-heavy pentaquarks}

Based on symmetry, there are multiple color-flavor configurations for the triply-heavy pentaquark states. The pentaquark state $nnQQ\bar{Q}$
that we investigate includes one heavy quark as an antiquark (where $n = u,d$ and $Q=c,b$), while the other four quarks are all valence quarks. We focus only on the heavy quark pentaquark systems with the highest symmetry, explicitly $nncc\bar{c}/bb\bar{b}$, as well as those with strangeness $S = -1, -2$, such as $nscc\bar{c}/bb\bar{b}$ and $sscc\bar{c}/bb\bar{b}$.

The color structure of pentaquarks can be viewed as the coupling between two substructures: one substructure consisting of three valence quarks and another substructure composed of a quark and an antiquark \cite{Weng:2019ynv, An:2020vku, An:2020jix, An:2021vwi, An:2022fvs}. By separately considering the color configurations of each substructure, two types of color singlet configurations can be obtained: one corresponds to a basis vector $\phi_{3}^{P}$
formed by the coupling of two color singlets $1 \otimes 1$, while the other involves two basis vectors, $\phi_{1}^{P}$ and $\phi_{2}^{P}$, resulting from the coupling of color octets $8 \otimes 8$. The specific forms of these three color basis vectors are provided in the appendix (\ref{apd:WF}) Eqs. (\ref{8b1}) to (\ref{8b3}). Since we only consider the ground state, the spin basis vectors for the pentaquarks can be constructed from the spins of the two substructures using Clebsch-Gordan coefficients, resulting in a total of 10 basis vectors, denoted as $\chi_{i}$ (where $i=1,2,3...10$), as shown in Eq. (\ref{Spin}). The color-spin basis vectors $\phi_{j}^{P} \chi_{i}^{P}$ can be derived based on the group algebra of $SU(3)_{c} \otimes SU(2)_{s}$, as presented in Eqs. (\ref{cs1}) to (\ref{cs3}). 

For pentaquark states, there are multiple flavor combination representations, such as $nscc\bar{c}$, which include $csn \otimes c\bar{c}$, $ccn \otimes s\bar{c}$, and $ccs \otimes n\bar{c}$. The choice of flavor symmetry basis vectors may vary; however, the eigenvalues of mass and magnetic moment remain consistent. Here, we consider the basis vectors for each flavor combination to facilitate the examination of the width ratios of the various final states. A summary of all the basis vectors is provided in Table \ref{tab:basis}.
 
According to the color-spin basis states of the pentaquarks given in Table \ref{tab:basis}, we can solve for the eigenvalues and eigenstates of Eq. (\ref{M}). The general form of the eigenstates is as follows: 
\begin{equation} 
 \psi = c_{1}\phi_{1}^{P} \chi_{1}^{P} +... c_{2}\phi_{2}^{P} \chi_{2}^{P} +... c_{3}\phi_{3}^{P} \chi_{1}^{P}+...,\label{equ:component}
\end{equation}
 where the eigenstate of the pentaquark system is a color-spin mixing state. The coefficient $c_{3}$ in front of $\phi_{3}^{P}$ describes the combination of color $1 \otimes 1$. When $|c_{3}|^2 > 0.8$, it is suggested that this configuration corresponds to scattering states of baryons and mesons. 

After obtaining the eigenstates of the hadrons, we can discuss the partial widths of the strong decays of the pentaquarks. Drawing on the methods employed in these studies \cite{Weng:2019ynv,Weng:2020jao,gaoc1992,Weng:2021ngd,An:2020vku}, we consider only the S-wave decay:
\begin{equation} 
\Gamma_i = \gamma_i\alpha k\cdot {|c_i|}^2. \label{equ:partialwidth}  
\end{equation}    
Here, $\Gamma_i$ represents the width of decay channel $i$, $\gamma_i$ is related to the decay dynamics, $\alpha$ is the coupling constant, $k$ is the momentum of the final state hadrons, and $|c_{i}|^2$ denotes the contribution of the corresponding decay final state component. For the two-body decay process $A \to B + C$, momentum conservation in the center-of-mass frame requires that the momentum of the final state products $k$ satisfies the equation 
\begin{equation}\label{phasespace}
m_A = \sqrt{m_B^2 + k^2} + \sqrt{m_C^2 + k^2},
\end{equation}
in the case of pentaquarks, $B$ represents a baryon and $C$ represents a meson. This decay mode is classified as OZI-superallowed \cite{Jaffe:1976ig}.

The initial state mass of the hadron can be obtained by solving Eq. (\ref{M}), while the final states can be determined based on flavor combinations and spin basis vectors. For example, for the pentaquark $nscc\bar{c}$ with total angular momentum $J = \frac{1}{2}$, there are several possible flavor combinations, such as $nsc \otimes c\bar{c}$, $ncc \otimes s\bar{c}$
, and $scc \otimes n\bar{c}$. The corresponding final state combinations of baryons and mesons should be $\Xi_{c}^{\ast} J/\psi$, $\Xi_{c}\eta_{c}$, $\Xi_{cc}^{\ast} \bar{D_{s}^{\ast}}$
, $\Xi_{cc} \bar{D_{s}}$,$\Omega_{cc} \bar{D}$ and $\Omega_{cc}^{\ast} \bar{D^{\ast}}$. In Eq. (\ref{equ:partialwidth}), the coefficients $c_i$ are determined by the coefficients of $\phi_{3}^{P}$
in the eigenstates. For certain decay final states whose baryon masses remain experimentally unknown in this paper, such as $\Xi_{cc}^{\ast}$, $\Xi_{bb}$, $\Xi_{bb}^{\ast}$, $\Omega_{cc}$
, $\Omega_{cc}^{\ast}$, and $\Omega_{bb}$, $\Omega_{bb}^{\ast}$, we use the results from the bag model calculations \cite{Zhang:2021yul}.
	
For specific processes, such as the decay final states with configuration $nsc \otimes c\bar{c}$
and $J = \frac{3}{2}$, possible states include $\Xi_{c}^{\ast} J/\psi$, $\Xi_{c}^{\ast} \eta_{c}$
, $\Xi_{c}^{0} J/\psi$, and $\Xi_{c}^{\prime} J/\psi$. In Eq. (\ref{phasespace}), the coefficients $\gamma_i$ depend on the spatial wave functions of the initial and final hadronic states, and in the heavy quark limit \cite{gaoc1992}, the values of $\gamma_i$ should satisfy the following relation:
\begin{equation}
\gamma_{\Xi_{c}^{\ast} J/\psi} = \gamma_{\Xi_{c}^{\ast} \eta_{c}} = \gamma_{\Xi_{c}^{0} J/\psi} = \gamma_{\Xi_{c}^{\prime} J/\psi}. \label{gamma}  
\end{equation}
Therefore, we can study the ratios of decay widths based on the factor $|c_i|^2 \cdot k$. Furthermore, $\Gamma_i$ strongly depends on $|c_i|^2 \cdot k$, which allows us to make rough estimates of the partial decay widths based on the values of $|c_i|^2 k$.

\label{sec:hadrons}
\subsection{$qqQQ\bar{Q}$ System}
\label{sec:qqQQQ}
	
\renewcommand{\tabcolsep}{0.4cm}
\renewcommand{\arraystretch}{0.9}
\begin{table*}[!htb]
	\caption{The bag radius $R$ of the pentaquark state $nncc\bar{c}$
		is measured in units of $\text{GeV}^{-1}$, with the representation of the eigenbasis given by $ccn \otimes n\bar{c}$, the mass spectrum is expressed in units of GeV, the magnetic moment is provided in units of $\mu_N$, and the binding energy $E_B$ is measured in units of GeV.}
		\label{table:nncccmass}
	\begin{tabular}{ccc|c|c|c|c}
		 \bottomrule[1.5pt]\bottomrule[0.5pt]
		$I$ &$J^{P}$ &$R_{0}$ & Eigenvector($ccn\otimes n\bar{c}$)&$M_{bag}$ &$\mu_{bag}$& $E_B$ \\ \hline
		$1$&${5/2}^{-}$ &5.571 &-  &5.837 &4.479,1.497,-1.484&-0.010    \\
		&${3/2}^{-}$    &5.570 &-0.14,-0.15,0.39,0.43,0.63,-0.30,-0.36  &5.963 &0.601,0.471,0.341&-0.010   \\
		&     &5.556 &0.72,-0.15,0.39,-0.36,0.01,0.19,-0.36  &5.831 &3.563,0.524,-2.515&-0.014  \\
		&    &5.476 &0.26,0.68,0.24,0.28,0.26,0.34,0.39  &5.784 &0.708,0.175,-0.358 &-0.032  \\
		&    &5,425 &-0.62,0.14,0.47,-0.40,-0.02,0.42,-0.17  &5.746 &2.654,0.410,-1.834&-0.044    \\
		&${1/2}^{-}$ &5.646&-0.11,0.20,0.23,0.22,-0.52,-0.72,-0.16,0.17&5.996 &0.963,0.753,0.544 &0.007 \\
		&     &5.562&0.41,-0.70,0.31,-0.30,-0.05,-0.11,-0.33,0.17&5.838 &3.731,1.157,-1.418 &-0.012 \\
		&     &5.450&0.20,-0.35,-0.03,0.71,0.23,-0.14,0.46,0.23 &5.767 &0.905,0.253,-0.399 &-0.038 \\
		&     &5.346& -0.17,0.30,0.54,-0.03,0.51,0.07,-0.13,0.55 &5.704 &0.744,0.300,-0.143 &-0.061\\
		$0$&   ${5/2}^{-}$  &5.595& - &5.913 &1.502 &-0.005\\
		&  ${3/2}^{-}$   &5.563&0,0.33,0.28,0.39,-0.68,-0.23,-0.37   &5.878 &1.346 &-0.012 \\
		&     &5.407&0,0.02,-0.49,0.32,0.07,0.61,-0.53 &5.781 &1.153&-0.048 \\
		&     &5.441&0,-0.6,0.32,0.43,-0.25,0.39,0.36&5.592 &0.110 &-0.040 \\
		&  ${1/2}^{-}$   &5.511&0.10,0.06,0.12,-0.53,0.23,-0.46,0.64,-0.14 &5.837 &0.301 &-0.024 \\
		&     &5.420&-0.13,-0.07,-0.10,0.17,0.58,-0.42,-0.44,-0.48 &5.755&-0.093 &-0.045\\
		&     &5.469&-0.80,-0.46,0.28,-0.01,-0.11,0.10,0.18,-0.15&5.603 &-0.804 &-0.034 \\
		&     &5.265&-0.3,-0.17,-0.67,-0.23,0.11,-0.21,-0.08,0.55&5.529 &0.004&-0.079\\
		 \bottomrule[1.5pt]\bottomrule[0.5pt]
	\end{tabular}
\end{table*}

\renewcommand{\tabcolsep}{0.15cm}
\renewcommand{\arraystretch}{0.9}
\begin{table*}[!htbp]
	\caption{The final states with a decay width ratio of 1 are assumed to have the largest $|c_i|^2 \cdot k$ values, with the corresponding final states having $|c_i|^2 \cdot k$ values of approximately 0.1 GeV denoted in parentheses. A “0” indicates a branching ratio less than $1 \times 10^{-4}$, the star denotes states below the threshold, and “\text{-}” represents scattering states.}
	\label{table:decaynnccc}
	\begin{tabular}{ccc|cccccc|cccc}
		\bottomrule[1.5pt]\bottomrule[0.5pt]
		&$ $&$ $ &$nnc \otimes c\bar{c}$ &$ $ &$ $ &$$&$$&$$&$ccn\otimes \bar{c}n$ &$ $ &$ $ &$ $\\  \hline
		$I$&$J^{P}$ &Mass &$\Lambda_{c}\eta_{c} $ &$\Lambda_{c}J/\psi$ &$\Sigma_{c}\eta_{c}$ &$\Sigma_{c}J/\psi$ &$\Sigma_{c}^{\ast}\eta_{c}$
		&$\Sigma_{c}^{\ast}J/\psi$ &$\Xi_{cc}D$&$\Xi_{cc}^{\ast}\bar{D^{\ast}}$&$\Xi_{cc}\bar{D^{\ast}}$&$\Xi_{cc}^{\ast}\bar{D}$
		\\ \hline
		$1$&${{5/2}^{-}}$
		&5.837  &$ $&$ $&$ $&$ $&$ $&\text{-}&$ $&$ $&$ $&$ $ \\ 	
		$ $&${{3/2}^{-}}$
		&5.963  &$ $&$$&$ $ &$0.006 $&$1(0.053)$&$0.014$&&$1$&$0.378$&$0.720$ \\ 
		&$ $&5.831  &$ $&$ $&$ $&$0.006$&$ 0.271$&$1$&$ $&$0$&$1(0.096)$&$0.292$   \\ 
		&$ $&5.784  &$ $&$$ &$ $&$ 1(0.121)$&$0.868 $ &$0.364 $&$ $&$0.283$&$1(0.097)$&$0.826$   \\
		&$ $&5.746&$ $&$ $&    & 1 & 0.623&0 &$ $&$0.001$&$0.017$&$1$(0.116)\\
  	    &${{1/2}^{-}}$
    	&5.996  &$ $&$ $&$0.024 $&$0.086$&$ $&$1(0.087) $&$0.078$&$1$&$0.056$&$ $ \\ 
	    &&5.838  &$ $&$ $&$0.029 $&$0.224 $&$ $&$1$&$0.321$&$0.083$&$1(0.081)$&$ $   \\ 
	    &&5.767  &$ $&$ $&$1 $&$0.315 $&$ $&$0.127 $&$0.365$&$0.051$&$1$&$ $   \\
	    &&5.704  &$ $&$ $&$1(0.139) $&$0.827 $&$ $&$0.009 $&$1$&$\ast$&$0.035$&$ $\\
		$0$&${{5/2}^{-}}$
	    &5.914  &$ $&$ $&$ $&$ $&$ $&$ $&$ $&\text{-}&$ $&$ $ \\ 	
		&${{3/2}^{-}}$
	    &5.878  &$ $&$1(0.013)$&$ $&$ $&$ $&$ $&$ $&$1 $&$0.380 $&$0.158 $ \\ 
	    &&5.781  &$ $&$1(0.008) $&$ $&$ $&$ $&$ $&$ $&$0.008$&$0.670$&$1$   \\ 
	    &&5.592  &$ $&$1(0.139) $&$ $&$ $&$ $&$ $&$ $&$\ast$&$\ast$&$1(0.024)$   \\
    	&${{1/2}^{-}}$
        &5.837  &$ 0.184$&$1(0.024) $&$ $&$ $&$ $&$ $&$0.066$&$0.036$&$1$&$ $ \\ 
        &&5.755  &$1(0.076) $&$0.001 $&$ $&$ $&$ $&$ $&$1$&$0.008$&$0.576$&$ $   \\ 
        &&5.603  &$0.072 $&$1 $&$ $&$ $&$ $&$ $&$1(0.012)$&$\ast$&$\ast$&$ $   \\
        &&5.529  &$ 1(0.144)$&$ 0.001$&$ $&$ $&$ $&$ $&$1(0.097)$&$\ast$&$\ast$&$ $\\
		\bottomrule[0.5pt]\bottomrule[1.5pt]
	\end{tabular}
\end{table*}

Next, we discuss the pentaquars $nncc\bar{c}$ with numerical results presented in Table \ref{table:nncccmass}. Due to spin symmetry, the states with $J^P = 5/2^{-}$
are all scattering states. For the isospin state with quantum number $I = 1$, the masses range in $5.70-6.00 \, \text{GeV}$, which exceeds the threshold of $\Sigma_{c}J/\psi$. For the isospin state with quantum number $I = 0$, the masses range in $5.53-5.92 \, \text{GeV}$, which exceeds the threshold of $\Lambda_{c}J/\psi$. Additionally, except for the state $(I, J^P, M) = (1, 1/2^{-}, 5.996)$, all other states have binding energies greater than $-10 \, \text{MeV}$.

In Table \ref{table:decaynnccc}, we present the decay branching ratios for the two color-flavor configurations of $nncc\bar{c}$, specifically $nnc \otimes c\bar{c}$
and $ccn \otimes \bar{c}n$. According to Eq. (\ref{gamma}), for spins determined by the flavor configuration of the final state, such as $(I, J^P) = (1,  3/2^{-})$
, we have $\gamma_{\Sigma_{c}J/\psi} = \gamma_{\Sigma_{c}^{\ast}\eta_{c}} = \gamma_{\Sigma_{c}^{\ast}J/\psi}$. Therefore, we only need to compute the values of $|c_i|^2 \cdot k$
and then calculate the ratios to obtain the partial decay width ratios. Here, we define the decay final state corresponding to the maximum value of $k \cdot |c_i|^2$
as 1 and indicate the final states with corresponding values less than $150\, \text{MeV}$ in parentheses. 

For the $nnc \otimes c\bar{c}$ configuration, the maximum values of $|c_i|^2 \cdot k$
for the $I=1$ states, such as $\left(3/2^{-}, 5.784\right)$ and $\left(1/2^{-}, 5.704\right)$, are $121\, \text{MeV}$, and $139\, \text{MeV}$, respectively. For the $I=0$ states $\left(3/2^{-}, 5.592\right)$, $\left(1/2^{-}, 5.837\right)$, $\left(1/2^{-}, 5.755\right)$, and $\left(1/2^{-}, 5.529\right)$, the corresponding maximum values of $|c_i|^2 \cdot k$
are $137\, \text{MeV}$, $24\, \text{MeV}$, $76\, \text{MeV}$ and $144\, \text{MeV}$, respectively.

Based on the results from Tables \ref{table:nncccmass} and \ref{table:decaynnccc}, the following states are noteworthy: $(I, J^P, M)$ = $(1, 3/2^{-}, 5.784)$, $(1, 1/2^{-}, 5.704)$, $(0, 3/2^{-}, 5.781)$, $(0, 1/2^{-}, 5.755)$, $(0, 3/2^{-}, 5.592)$, and $(0, 1/2^{-}, 5.529)$. On one hand, these states have binding energies exceeding -30 MeV; on the other hand, their decay widths are suppressed due to the relatively small values of $|c_i|^2 \cdot k$. Therefore, it is very likely that they will be observed experimentally through decay channels involving $\Sigma_{c} J/\psi$ and $J/\psi \Lambda_{Q}$.

\renewcommand{\tabcolsep}{0.4cm}
\renewcommand{\arraystretch}{0.9}
\begin{table*}[!htb]
	\caption{The bag radius $R$ of the pentaquark state $nnbb\bar{b}$
		is measured in units of $\text{GeV}^{-1}$, with the representation of the eigenbasis given by $bb\otimes n\bar{b}$, the mass spectrum is expressed in units of GeV, the magnetic moment is given in units of $\mu_N$, and the binding energy $E_B$ is measured in units of GeV.}
	\label{table:massnnbbb}
	\begin{tabular}{ccc|c|c|c|c}
		\bottomrule[1.5pt]\bottomrule[0.5pt]
		$I$&$J^{P}$ &$R_{0}$ &Eigenvector($bb\otimes n\bar{b}$)&$M_{bag}$ &$\mu_{bag}$& $E_B $ \\ \hline
		&${5/2}^{-}$    &5.013 &- &15.882 &3.491,0.808,-1.874  &-0.134\\
		&${3/2}^{-}$    &5.004 &0.04,0.31,-0.33,-0.36,-0.57,0.37,0.45  &16.090 &-0.071,-0.081,-0.091  &-0.136\\
		&     &5.009 &-0.70,0.32,-0.18,0.45,0.18,-0.09,0.37 &15.883 &3.497,0.935,-1.628&-0.135\\
		&    &4.977 &0.30,0.57,0.52,0.11,0.34,0.42,0.13&15.866 &0.788,0.201,-0.387 &-0.141\\
		&    &4.943 &-0.65,-0.07,0.41,-0.46,-0.07,0.31,-0.30  &15.849 &3.037,0.811,-1.416&-0.149\\
		&${1/2}^{-}$ &5.023&-0.03,0.05,0.39,0.18,-0.39,-0.71,-0.22,0.33   &16.095 &-0.127,-0.136,-0.146&-0.132\\
		&     &5.019&0.42,-0.73,0.28,-0.30,0.16,0.09,-0.25,0.18  &15.887 &3.262,0.785,-1.692  &-0.133 \\
		&     &4.961&-0.18,0.31,0.10,-0.72,-0.23,0.10,-0.50,-0.18 &15.859 &0.644,0.160,-0.325&-0.145  \\
		&     &4.917& 0.20,-0.35,-0.45,-0.01,-0.59,-0.19,0.04,-0.49 &15.839 &0.703,0.163,-0.377&-0.154 \\
		0&   ${5/2}^{-}$  &5.014&- &16.012 &0.809&-0.134  \\
		&  ${3/2}^{-}$   &4.981&0,-0.40,-0.21,-0.35,0.64,0.25,0.44&15.994 &0.809  &-0.154\\
		&     &4.907&0,-0.04,-0.47,0.33,0.08,0.65,-0.49&15.959 &0.272 &-0.156\\
		&     &4.902&0,-0.55,0.38,0.46,-0.33,0.32,0.35&15.660 &-0.002 &-0.157\\
		&  ${1/2}^{-}$   &4.947&0.03,0.02,0.21,-0.52,0.13,-0.37,0.71,-0.13  &15.977 &0.645 &-0.148 \\
		&     &4.869&0.03,0.02,-0.15,-0.18,-0.53,0.37,0.31,0.65 &15.937&-0.343&-0.164 \\
		&     &4,910&-0.79,-0.45,0.29,0.07,-0.16,0.17,0.10,-0.14&15.662 &0.138  &-0.156 \\
		&     &4.872&-0.36,-0.21,-0.64,-0.22,0.30,-0.37,-0.14,0.35&15.653&0.017  &-0.164 \\
		\bottomrule[0.5pt]\bottomrule[1.5pt]
	\end{tabular}
\end{table*}

\renewcommand{\tabcolsep}{0.25cm}
\renewcommand{\arraystretch}{0.9}
\begin{table*}[!t]
\caption{The final states with a decay width ratio of 1 are assumed to have the largest $|c_i|^2 \cdot k$ values, with the corresponding final states having $|c_i|^2 \cdot k$ values of approximately 0.1 GeV denoted in parentheses. A “0” indicates a branching ratio less than $1 \times 10^{-4}$, the star denotes states below the threshold, and “\text{-}” represents scattering states.}
	\label{table:decaynnbbb}
	\begin{tabular}{ccc|cccccc|cccc}
		\bottomrule[1.5pt]\bottomrule[0.5pt]
		&$ $&$ $ &$nnb \otimes b\bar{b}$ &$ $ &$ $ &$$&$$&$$&$bbn\otimes \bar{b}n$ &$ $ &$ $ &$ $\\  \hline 
		$I$&$J^{P}$ &Mass &$\Lambda_{b}\eta_{b} $ &$\Lambda_{b}\Upsilon$ &$\Sigma_{b}\eta_{b}$ &$\Sigma_{b}\Upsilon$ &$\Sigma_{b}^{\ast}\eta_{b}$
		&$\Sigma_{b}^{\ast}\Upsilon$ &$\Xi_{bb}\bar{B}$&$\Xi_{bb}^{\ast}\bar{B^{\ast}}$&$\Xi_{bb}\bar{B^{\ast}}$&$\Xi_{bb}^{\ast}\bar{B}$
		\\ \hline
		$1$&${{5/2}^{-}}$
		&15.882  &$ $&$ $&$ $&$ $&$ $&\text{-}&$ $&$ $&$ $&$ $ \\ 	
		$ $&${{3/2}^{-}}$
		&16.090  &$ $&$ $&$ $&$ 0$&$1(0.011)$&$0$&$$ &$1$&$0.662$&$0.461$ \\ 
		&$ $&15.883&$ $&$ $&$ $  &$0 $ &$0.783 $&$ 1$&$ $&$0.204$&$1$&$0.056$   \\ 
		&$ $&15.866  &$$&$$&$ $&$0.288$&$0.654 $ &$1$&$ $ &$0.595$&$0.1$&$1$  \\
		&$ $&15.849  &$ $&$ $&  & 1 &0.220 &0.005 &$ $&$0.045$&$0.983$&$1$\\
		&${{1/2}^{-}}$
		&16.095  &$ $&$ $&$0.063 $&$0.303 $&$1(0.013) $&$ $&$0.239$&$1$&$0.103$&$ $ \\ 
		&&15.887  &$ $&$ $&$0.014 $&$0.071 $&$1 $&$ $&$0.573$&$0.125$&$1(0.082)$&$ $   \\ 
		&&15.859  &$ $&$ $&$ 1$&$0.363 $&$0.100 $&$ $&$0.137$&$0.032$&$1$&$ $   \\
		&&15.839  &$ $&$ $&$ 0.727$&$1 $&$0 $&$ $&$1$&$0.280$&$0.014$&$ $\\
		$0$&${{5/2}^{-}}$
		&16.012  &$ $&$ $&$ $&$ $&$ $&$ $&$ $&\text{-}&$ $&$ $ \\ 	
		&${{3/2}^{-}}$
		&15.994  &$ $&$1(0.014) $&$ $&$ $&$ $&$ $&$ $&$1 $&$0.163 $&$0.505 $ \\ 
		&&15.959  &$ $&$1(0.012) $&$ $&$ $&$ $&$ $&$ $&$0.016$&$1$&$0.581$   \\ 
		&&15.660  &$ $&$1 $&$ $&$ $&$ $&$ $&$ $&$\ast$&$0.769$&$1(0.051)$   \\
		&${{1/2}^{-}}$
		&15.977 &$ 0.125$&$1(0.004) $&$ $&$ $&$ $&$ $&$0.035$&$0.25$&$1$&$ $ \\ 
		&&15.937  &$1(0.014) $&$0 $&$ $&$ $&$ $&$ $&$1$&$0.277$&$0.216$&$ $   \\ 
		&&15.662  &$0.050 $&$1 $&$ $&$ $&$ $&$ $&$1(0.013)$&$\ast$&$0.346$&$ $   \\
		&&15.653  &$ 1$&$ 0.003$&$ $&$ $&$ $&$ $&$1(0.081)$&$\ast$&$0.085$&$ $\\
		
		\bottomrule[0.5pt]\bottomrule[1.5pt]
	\end{tabular}
\end{table*}

The results for the $nnbb\bar{b}$ present in Table \ref{table:massnnbbb}. The masses of this system range in $15.60-16.10 \, \text{GeV}$, exhibiting smaller mass splittings, magnetic moments, and radii compared to the $nncc\bar{c}$ system, reflecting the suppressive effect of the heavy quark cluster. Additionally, the increase in bag binding energy is particularly notable, with a binding energy $E_B$ of approximately 150 MeV, indicating a higher degree of compact stability for the $nnbb\bar{b}$
system compared to $nncc\bar{c}$. Furthermore, in Table \ref{table:decaynnbbb}, we present the decay branching ratios for the color-flavor configurations $nnb \otimes b\bar{b}$
and $nbb \otimes n\bar{b}$. The states with quantum numbers $(I, J^P, M)$ are $(1, 5/2^{-}, 15.882)$ and $(0, 5/2^{-}, 16.012)$, which are believed to correspond to the scattering states of $\Sigma_{b}^{\ast}\Upsilon$ and $\Xi_{bb}^{\ast}B^{\ast}$, respectively. For the $nnb \otimes b\bar{b}$
configuration, the corresponding quantum numbers are $(1, 3/2^{-}, 16.090)$, $(1,  1/2^{-}, 16.095)$, $(0, 3/2^{-}, 15.994)$, $(0, 3/2^{-}, 15.959)$
, $(0, 1/2^{-}, 15.977)$, and $(0, 1/2^{-}, 15.937)$. The strong decay widths of these states are suppressed by the reduced values of $|c_i|^2 \cdot k$.

\renewcommand{\tabcolsep}{0.4cm}
\renewcommand{\arraystretch}{0.9}
\begin{table*}[!htb]
	\caption{The bag radius $R$ of the pentaquark state $nscc\bar{c}$
		is measured in units of $\text{GeV}^{-1}$, with the representation of the eigenbasis given by $ccn\otimes s\bar{c}$, the mass spectrum is expressed in units of GeV, the magnetic moment is given in units of $\mu_N$, and the binding energy $E_B$ is measured in units of GeV.}
	\label{table:massnsccc}	
	\begin{tabular}{cccccc}
		\bottomrule[1.2pt]\bottomrule[0.5pt]
		$J^{P}$ &$R_{0}$&Eigenvector($ccn\otimes s\bar{c}$)&$M_{bag}$ &$\mu_{bag}$&$E_B $ \\ \hline
	    ${5/2}^{-}$ &$5.557$&-&$5.963$ &$1.733,-1.219$&-0.013  \\
		    &5.582& - &6.015 &1.732,-1.219&-0.008  \\
		${3/2}^{-}$     &5.421&0.03,-0.62,0.27,0.45,-0.22,0.39,0.37  &5.764 &0.230,-0.038&-0.044 \\
		    &5.527&0.54,-0.21,-0.64,0.41,0.05,-0.24,-0.14 &5.878 &0.192,-0.481 & -0.020\\
		    &5.542&0.33,0.18,0.27,-0.01,0.03,-0.55,0.70 &5.896 &1.354,-0.105 &-0.017\\
		    &5.546&0.22,0.64,-0.01,0.38,0.25,0.55,0.16&5.916 &1.395,-0.245 &-0.016\\
		    &5.548&0.71,-0.07,0.49,-0.26,-0.09,0.14,-0.40 &5.956 &0.718,-1.132 &-0.015 \\
		    &5.556&-0.12,0.33,0.20,0.47,-0.68,-0.27,-0.28&5.986&2.965,-2.061 &-0.014 \\
		    &5.583&0.17,0.11,-0.41,-0.44,-0.64,0.30,0.31&6.052 &1.150,0.849&-0.007\\
		${1/2}^{-}$ &5.314&-0.28,-0.16,-0.67,-0.25,0.05,-0.17,-0.06,0.58  &5.702 &0.303,0.496&-0.068  \\
		     &5.384&0.79,0.47,-0.27,0.04,0.11,-0.07,-0.19,0.17&5.775 &-0.270,-0.560&-0.053 \\
		     &5.432&0.21,-0.27,-0.54,0.03,-0.53,-0.01,0.18,-0.52 &5.842 &0.475,0.144 &-0.042 \\
		     &5.452&0.18,0.03,0.19,0.01,-0.44,0.36,0.57,0.52&5.879 &-0.278,0.884&-0.039 \\
		     &5.485&-0.16,0.38,0.06,-0.67,-0.37,0.30,-0.37,-0.10&5.899 &0.164,-0.017 &-0.030\\
		     &5.518&0.08,-0.37,0.05,0.42,-0.26,0.37,-0.65,0.22&5.946 &0.002,-1.982 &-0.022 \\
		     &5.516&0.41,-0.58,0.32,-0.52,-0.01,-0.33,-0.08,0.09&5.965 &1.581,-1.041&-0.023 \\
		     &5.593&-0.14,0.25,0.19,0.22,-0.54,-0.71,-0.13,0.14&6.086 &1.004,1.121 &-0.005\\
		\bottomrule[0.5pt]\bottomrule[1.5pt]
	\end{tabular}
\end{table*}

\renewcommand{\tabcolsep}{0.06cm}
\renewcommand{\arraystretch}{0.9}
\begin{table*}[!htbp]
	\caption{The final states with a decay width ratio of 1 are assumed to have the largest $|c_i|^2 \cdot k$ values, with the corresponding final states having $|c_i|^2 \cdot k$ values of approximately 0.1 GeV denoted in parentheses. A “0” indicates a branching ratio less than $1 \times 10^{-4}$, the star denotes states below the threshold, and “\text{-}” represents scattering states.}
	\label{table:decaynsccc}	
	\begin{tabular}{cc|cccc|cccc|cccccc}
	\bottomrule[1.2pt]\bottomrule[0.5pt]
		&$ $ &$ccn\otimes s\bar{c}$ &$ $ &$ $ &$$&$ccs\otimes n\bar{c}$&$ $& & &$cns\otimes c\bar{c}$ & &$ $ &$ $ &$ $&\\ 
		$J^{P}$ &Mass &$\Xi_{cc}^{\ast}\bar{D_{s}^{\ast}}$ &$\Xi_{cc}\bar{D_{s}^{\ast}}$ &$\Xi_{cc}^{\ast}\bar{D_{s}}$ &$\Xi_{cc}\bar{D_{s}}$ &$\Omega_{cc}^{\ast}\bar{D^{\ast}}$
		&$\Omega_{cc}\bar{D^{\ast}}$ &$\Omega_{cc}^{\ast}\bar{D}$&$\Omega_{cc}\bar{D}$&
		$\Xi_{c}^{\ast} J/\psi$&$\Xi_{c}^{\ast}\eta_{c}$&$\Xi_{c}^{0}\eta_{c}$&$\Xi_{c}^{\prime}\eta_{c}$&$\Xi_{c}^{0}J/\psi$&$\Xi_{c}^{\prime}J/\psi$
		\\ \hline
		${{5/2}^{-}}$
		&5.963  &$ $&$ $&$ $&$ $&$ $&$ $&$ $&$ $&\text{-} &$ $&&&& \\
		&6.015  &\text{-}&$ $&$ $&$ $&$ $&$ $&$ $&$ $&$ $&$ $&&&& \\ 	
		${{3/2}^{-}}$
		&5.764  &$\ast $&$0.558$&$1(0.071)$&$$&$\ast$&$0.396$&$1(0.085)$&$$&$0$&$0.002$&&&0.393&1(0.124) \\ 
		&5.878  &$0.026 $&$ 0.277$&$1(0.043)$&$$&$0.003 $&$0.169$&$1$&$$&$0.003$&$0.980$&&&1&0.148   \\ 
		&5.896  &$0.001$&$1$ &$0.697 $&$$&$0.042$&$0.187$&$0.159$&$ $&$0.039$&$0.012$& &&$1(0.081)$ & $0.449$   \\
		&5.916  &$0.133$&$0.075$&$1$& &$0.070$&$1$&$0.014$&$$&$0.070$&$1$&&&0.020&0.014\\
		&5.956  &$0.042$&$1(0.124)$&$0.127$& &$0.269$&1&$0.762$&$$&$1$&$0.217$&&&0.005&0.009\\
		&5.986  &$1$&$0.207$&$0.211$& &$1$&$0.510$&$0.093$&$$&$0.305$&$0.203$&&&0.360&$1(0.011)$\\
		&6.052  &$1$&$0.275$&$0.279$& &$1$&$0.352$&$0.200$&&$0.859$&$1(0.068)$&&&0.006&0.003\\
		${{1/2}^{-}}$
		&5.702&$\ast$&$\ast$&$ $&$1$&$\ast$&$\ast$&$$&$1$&$\ast$&$$&0.0434&1(0.134)&0&0 \\ 
		&5.775  &$\ast$&$0.65$&$ $&$1(0.020)$&$\ast$&$1(0.020)$&$$&$0.923$&$0$&&$0.049$&0.108&0.423&1   \\ 
		&5.842  &$0$&$0.082$&$ $&$1$&$0$&$0.025$&$$&$1$&$0.014$&$$&$0.976$&$0.428$&$1(0.116)$&0.208   \\
		&5.879  &$0.238$&$0.894$&$$&$1$&$0.765$&$0.968$&$$&$1$&$0.021$&$ $&0.053&$1(0.079)$&0.134&0.030\\
		&5.899  &$0.438 $&$1(0.093) $&$ $&$ 0.105$&$0 $&$0.64 $&$$&$1$&$0.125$&$ $&1&0.103&0.233&0.072 \\ 
		&5.946  &$0.235$&$ 1$&$ $&$0.149 $&$0.987 $&$1 $&$$&$0.014$&$1(0.057)$&$ $&0.031&0.192&0&0.504   \\ 
		&5.965  &$1(0.066) $&$ 0.072$&$ $&$0.117 $&0.006&1&&0.124&1&$ $&0.023&0.001&0.359&0.024   \\
		&6.086  &$ 1$&$ 0.041$&$ $&$0.057 $&$1 $&$0.048 $&$$&$0.051$&$1(0.113)$&&0.011&0.004&0.051&0.014\\
		
		\bottomrule[0.5pt]\bottomrule[1.5pt]
	\end{tabular}
	\label{table:tab2}
\end{table*}
For the system $nscc\bar{c}$ with strangeness $S = -1$, the results in Table \ref{table:massnsccc} indicate that the mass range of these states is $6.10-5.70 \text{GeV}$. For most states, the binding energy is greater than $-10 \text{MeV}$, with some states even reaching $-40 \text{MeV}$. We present the decay branching ratios for three color-flavor configurations of the $nscc\bar{c}$ state: $nsc \otimes c\bar{c}$, $ccn \otimes \bar{c}s$, and $ccs \otimes \bar{c}n$, as shown in Table \ref{table:decaynsccc}. Our specific focus is on the decay modes of the $nsc \otimes c\bar{c}$configuration, particularly the final states $\Xi_{c}^{\ast} J/\psi$
, $\Xi_{c}^{\ast} \eta_{c}$, $\Xi_{c}^{0} \eta_{c}$, $\Xi_{c}^{\prime} \eta_{c}$, $\Xi_{c}^{0} J/\psi$, and $\Xi_{c}^{\prime} J/\psi$, for which we provide the corresponding branching ratios. These states, such as $\left( 3/2^{-}, 5.764 \right)$, $\left(1/2^{-}, 5.702 \right)$, and $\left( 1/2^{-}, 5.842 \right)$, exhibit significant compact stability, while their decays are suppressed by the factor $|c_i|^2 \cdot k$.

\renewcommand{\tabcolsep}{0.4cm}
\renewcommand{\arraystretch}{1}
\begin{table*}[!htbp]
	\caption{The bag radius $R$ of the pentaquark state $nsbb\bar{b}$
		is measured in units of $\text{GeV}^{-1}$, with the representation of the eigenbasis given by $bbn\otimes s\bar{b}$, the mass spectrum is expressed in units of GeV, the magnetic moment is given in units of $\mu_N$, and the binding energy $E_B$ is measured in units of GeV.}
	\label{table:massnsbbb}	
	\begin{tabular}{cccccc}
		\bottomrule[1.5pt]\bottomrule[0.5pt]
		$J^{P}$ &$R_{0}$ &Eigenvector($bbn\otimes s\bar{b}$)&$M_{bag}$ &$\mu_{bag}$&$E_B$\\ \hline
		${5/2}^{-}$     &5.054& -&16.001 &1.012,-1.668&-0.125  \\
		$ $     &5.061&- &16.097 &1.012,-1.668&-0.124 \\
		${3/2}^{-}$     &4.935&0.01,-0.56,0.37,0.46,-0.33,0.32,0.35   &15.821 &0.001,-0.023&-0.150 \\
		$ $     &4.925&-0.65,-0.06,0.43,-0.45,-0.08,0.31,-0.29  &15.971 &0.484,-0.659 &-0.152\\
		$ $     &4.987&0.29,0.59,0.53,0.11,0.33,0.39,0.17  &15.986 &0.111,-0.244 & -0.193\\
		$ $     &5.049&-0.70,0.30,-0.20,0.43,0.17,-0.11,0.39 &16.002 &0.632,-0.749&-0.126\\
		$ $     &5.084&-0.02,-0.01,-0.44,0.35,0.11,0.65,-0.50&16.050 &0.679,-1.581 &-0.119\\
		$ $     &5.092&0.01,-0.40,-0.21,-0.37,0.64,0.26,0.42&16.081 &1.872,-2.925&-0.117\\
		$ $     &5.144&-0.05,-0.30,0.35,0.36,0.57,-0.38,-0.43  &16.157 &0.094,-0.125 &-0.106\\
		${1/2}^{-}$     &4.896&-0.35,0.20,-0.66,-0.23,-.29,-0.36,-0.14,0.37&15.814 &0.041,0.011 &-0.159\\
		$ $     &4.953&0.79,0.46,-0.29,-0.06,0.16,-0.16,-0.10,0.14&15.823 &0.112,0.059  &-0.147\\
		$ $     &4.982&0.20,-0.34,-0.46,-0.01,-0.60,-0.17,0.05,-0.49&15.962 &0.151,-0.178 &-0.140\\
		$ $     &5.012&-0.18,0.32,0.09,-0.70,-0.22,0.10,-0.52,-0.19&15.980 &0.095,-0.153 &-0.134\\
		$ $     &5.074&0.42,-0.73,0.27,-0.28,0.15,0.08,-0.28,0.18&16.005 &0.532,-0.780  &-0.121\\
		$ $     &5.085&0.04,0.03,-0.12,-0.20,-0.53,0.39,0.31,0.64&16.030 &-0.812,0.827 &-0.118\\
		$ $     &5.104&0.04,0.01,0.21,-0.55,0.13,-0.37,0.69,-0.13&16.065 &1.448,-1.86&-0.114\\
		$ $     &5.198&0.03,-0.06,-0.38,-0.18,0.41,0.71,0.21,-0.32&16.162 &-0.332,-0.266&-0.094\\
		\bottomrule[0.5pt]\bottomrule[1.5pt]
	\end{tabular}
\end{table*}
\renewcommand{\tabcolsep}{0.05cm}
\renewcommand{\arraystretch}{1}
\begin{table*}[!htbp]
	\caption{The final states with a decay width ratio of 1 are assumed to have the largest $|c_i|^2 \cdot k$ values, with the corresponding final states having $|c_i|^2 \cdot k$ values of approximately 0.1 GeV denoted in parentheses. A “0” indicates a branching ratio less than $1 \times 10^{-4}$, the star denotes states below the threshold, and “\text{-}” represents scattering states.}
	\label{table:decaynsbbb}	
	\begin{tabular}{cc|cccc|cccc|cccccc}
		\bottomrule[1.5pt]\bottomrule[0.5pt]
		&$ $ &$bbn\otimes s\bar{b}$ &$ $ &$ $ &$$&$bbs\otimes n\bar{b}$&$ $& & &$bns\otimes b\bar{b}$ & &$ $ &$ $ &$ $&\\ 
		$J^{P}$ &Mass &$\Xi_{bb}^{\ast}\bar{B_{s}^{\ast}}$ &$\Xi_{bb}^{\ast}\bar{B_{s}}$ &$\Xi_{bb}\bar{B_{s}^{\ast}}$ &$\Xi_{bb}\bar{B_{s}}$ &$\Omega_{bb}^{\ast}\bar{B^{\ast}}$
		&$\Omega_{bb}^{\ast}\bar{B^{\ast}}$ &$\Omega_{bb}\bar{B^{\ast}}$&$\Omega_{bb}\bar{B}$&
		$\Xi_{b}^{\ast} \Upsilon$&$\Xi_{b}^{\ast}\eta_{b}$&$\Xi_{b}^{0}\eta_{b}$&$\Xi_{b}^{\prime}\eta_{b}$&$\Xi_{b}^{0}\Upsilon$&$\Xi_{b}^{\prime}\Upsilon$
		\\ \hline
		${{5/2}^{-}}$
		&16.001  &$ $&$ $&$ $&$ $&$ $&$ $&$ $&$ $&\text{-}&$ $&&&&\\
		&16.097 &\text{-}&$ $&$ $&$ $&$ $&$ $&$ $&$ $&$ $&$ $&&&&\\ 	
		${{3/2}^{-}}$
		&15.821  &0.842&0.819&1&&0.629&0.911&1&&0&0&&&0.361&1\\ 
		&15.971  &0.064&1&0.894&&0.025&1&0.891&&0.005&0.285&&&1&0.317\\ 
		&15.986  &0.716&1&0.192&&0.553&1&0.091&&1&0.729&&&0.285&0.083\\
		&16.002  &0.192&0.072&1&&0.159&0.061&1&&1&0.714&&&0&0\\
		&16.050  &0.031&1&0.573&&0.006&1&0.602&&0&1(0.007)&&&0.067&0.181\\
		&16.081  &1&0.164&0.441&&1&0.145&0.496&&0.240&0&&&0.580&1(0.005)\\
		&16.157 &$1$&$0.449$&$0.552$& &1&0.399&0.674&&0&1(0.015)&&&0&0\\
		${{1/2}^{-}}$
		&15.814&0.531&&0.121&1&0.545&&0.121&1&0&&0.386&1&0.001&0\\ 
		&15.823&0.809&&0.462&1&0.83&&0.462&1&0&&0.021&0.055&0.396&1\\ 
		&15.962&0.104&&0.009&1&0.105&&0.009&1&0&&0.748&0.216&1&0.313\\
		&15.980&0.033&&1&0.149&0.034&&1&0.150&0.126&&1&0.315&0.367&0.166\\
		&16.005&0.078&&1&0.503&0.079&&1&0.476&1&&0.0122&0.004&0.063&0.020 \\ 
		&16.030&0.309&&0.214&1&0.312&&0.214&1&0&&0.123&1(0.016)&0&0\\ 
		&16.065&0.255&&1&0.040&0.264&&1&0.040&0.154&&0.133&0.425&0.047&1(0.005) \\
		&16.162&1&&0.092&0.221&1&&0.091&0.219&1(0.024)&&0.113&0.020&0.219&0.069\\
		
			\bottomrule[0.5pt]\bottomrule[1.5pt]
	\end{tabular}
	\label{table:tab2}
\end{table*}

The numerical values for $nsbb\bar{b}$ are listed in Table \ref{table:massnsbbb}. The average bag radius is approximately $5.0 \, \text{GeV}^{-1}$, with a mass range of $16.15-15.81 \, \text{GeV}$. The mass splitting is very small, and the bag binding energy is about $-130 \, \text{MeV}$. Indicators such as the magnetic moment suggest a dependence on the heavy quarks. In Table \ref{table:decaynsbbb}, we present the decay width ratios for $nsbb\bar{b}$. We pay particular attention to the decay channels of the $nsb \otimes b\bar{b}$ configuration, with final states including $\Xi_{b}^{\ast} \Upsilon$, $\Xi_{b}^{\ast} \eta_{b}$, $\Xi_{b}^{0} \eta_{b}$, $\Xi_{b}^{\prime} \eta_{b}$, $\Xi_{b}^{0} \Upsilon$, and $\Xi_{b}^{\prime} \Upsilon$. The results indicate that the decay widths of the following states: $\left( 3/2^{-}, 16.050 \right)$, $\left( 3/2^{-}, 16.081 \right)$, $\left( 3/2^{-}, 16.157 \right)$, $\left( 1/2^{-}, 16.030 \right)$, $\left( 1/2^{-}, 16.065 \right)$, and $\left( 1/2^{-}, 16.162 \right)$, are suppressed by the factor $|c_i|^2 \cdot k$.

\renewcommand{\tabcolsep}{0.4cm}
\renewcommand{\arraystretch}{1}
\begin{table*}[!t]
	\caption{The bag radius $R$ of the pentaquark state $sscc\bar{c}$
		is measured in units of $\text{GeV}^{-1}$, with the representation of the eigenbasis given by $ccs\otimes s\bar{c}$, the mass spectrum is expressed in units of GeV, the magnetic moment is given in units of $\mu_N$, and the binding energy $E_B$ is measured in units of GeV.}
	\label{table:massssccc}	
	\begin{tabular}{cccccc}
	\bottomrule[1.5pt]\bottomrule[0.5pt]
		$J^{P}$ &$R_{0}$&Eigenvector($ccs\otimes s\bar{c}$) &$M_{bag}$ &$\mu_{bag}$&$E_{B} $  \\ \hline
		${5/2}^{-}$     &5.626&-  &6.093 &-0.993&0.003  \\
		${3/2}^{-}$     &5.643&0.21,0.04,-0.40,-0.48,-0.65,0.25,0.27   &6.147 &0.195&0.007  \\
		$ $     &5.614&0.71,-0.07,0.47,-0.30,0.10,0.22,-0.34 &6.087 &-1.961 &-0.001 \\
		$ $     &5.535&-0.32,-0.66,-0.04,-0.34,-0.18,-0.25,-0.50&6.043 &-0.379 &-0.018 \\
		$ $     &5.482&0.21,0.04,-0.40,-0.48,-0.66,0.25,0.27&6.014 &-1.255&-0.031\\
		${1/2}^{-}$     &5.719&-0.18,0.31,0.14,0.24,-0.55,-0.69,-0.11,0.09   &6.183 &-0.379 &0.024\\
		$ $     &5.608&-0.38,0.66,-0.30,0.31,0.18,0.22,0.37,-0.14 &6.088 &-0.785&-0.002\\
		$ $     &5.512&0.22,-0.39,-0.01,0.69,0.19,-0.18,0.43,0.24 &6.032 &-0.333&-0.024\\
		$ $     &5.414&-0.16.0.27,0.58,-0.02,0.46,0.01,-0.16,0.57&5.975 &0.014 &-0.046\\
		\bottomrule[0.5pt]\bottomrule[1.5pt]
	\end{tabular}
\end{table*}

\renewcommand{\tabcolsep}{0.4cm}
\renewcommand{\arraystretch}{1}
\begin{table*}[!t]
\caption{The final states with a decay width ratio of 1 are assumed to have the largest $|c_i|^2 \cdot k$ values, with the corresponding final states having $|c_i|^2 \cdot k$ values of approximately 0.1 GeV denoted in parentheses. A “0” indicates a branching ratio less than $1 \times 10^{-4}$, the star denotes states below the threshold, and “-” represents scattering states.}
	\label{table:decayssccc}	
	\begin{tabular}{cc|cccc|cccc}
		\bottomrule[1.5pt]\bottomrule[0.5pt]
		&$ $ &$ccs\otimes s\bar{c}$ &$ $ &$ $ &$$&$css\otimes c\bar{c}$&$ $& & \\
		$J^{P}$ &Mass &$\Omega_{cc}^{\ast}\bar{D_{s}^{\ast}}$ &$\Omega_{cc}\bar{D_{s}^{\ast}}$ &$\Omega_{cc}^{\ast}\bar{D_{s}}$ &$\Omega_{cc}\bar{D_{s}}$ &$\Omega_{c}^{\ast}J/\psi$
		&$\Omega_{c}^{\ast}\eta_{c}$ &$\Omega_{c}J/\psi$&$\Omega_{c}\eta_{c}$\\ \hline
		${{5/2}^{-}}$
		&6.093  &\text{-}&$ $&$ $&$ $&$ $&$ $&$ $&$ $\\
		${{3/2}^{-}}$
		&6.147  &1&0.207&0.185&&0.012&&0.034&1\\ 
		&6.087  &0.064&1(0.098)&0.423&&0.017&&1&0.134\\ 
		&6.043  &0.094&1&0.282&&1&&0.116&0.304\\
		&6.014  &0.001&0.008&1&&0.908&&0.002&1\\
		${{1/2}^{-}}$
		&6.183&1&0.028&&0.026&0.034&0.008&1&\\ 
		&6.088&0.283&1&&0.169&0.419&0.611&1&\\ 
		&6.032&0.121&1&&0.413&0.254&1&0.152&\\
		&5.975&0&&0.055&1&0.790&1&0.021&\\
		\bottomrule[0.5pt]\bottomrule[1.5pt]
	\end{tabular}
\end{table*}

\renewcommand{\tabcolsep}{0.5cm}
\renewcommand{\arraystretch}{1}
\begin{table*}[!htbp]
	\caption{The bag radius $R$ of the pentaquark state $ssbb\bar{b}$
		is measured in units of $\text{GeV}^{-1}$, with the representation of the eigenbasis given by $bbs\otimes s\bar{b}$, the mass spectrum is expressed in units of GeV, the magnetic moment is given in units of $\mu_N$, and the binding energy $E_B$ is measured in units of GeV.}
	\label{table:massssbbb}	
	\begin{tabular}{cccccc}
		\bottomrule[1.5pt]\bottomrule[0.5pt]
		$J^{P}$ &$R_{0}$&Eigenvector($bbs\otimes s\bar{b}$) &$M_{bag}$ &$\mu_{bag}$&$E_{B}$ \\ \hline
		${5/2}^{-}$     &5.103&-  &16.124 &-1.482&-0.115  \\
		${3/2}^{-}$     &5.106&-0.06,-0.27,0.35,0.38,0.58,-0.36,-0.43&16.229 &-0.996&-0.114\\
		$ $     &5.099&-0.7,0.30,-0.23,0.45,0.13,-0.10,0.38&16.124 &-1.280 &-0.115\\
		$ $     &5.069&0.28,0.60,0.48,0.13,0.33,0.42,0.18&16.109 &-0.278 &-0.122\\
		$ $     &5.042&0.65,0.04,-0.42,0.45,0.06,-0.33,0.29&16.095 &-1.089&-0.128\\
		${1/2}^{-}$     &5.127&-0.04,0.08,0.35,0.19,-0.43,-0.72,-0.21,0.29  &16.235 &-0.149 &-0.109\\
		$ $     &5.104&0.42,-0.73,0.29,-0.30,0.11,0.04,-0.27,0.18 &16.126 &-1.338&-0.114\\
		$ $     &5.057&0.19,-0.32,-0.09,0.72,0.22,-0.10,0.49,0.18 &16.104 &-0.272&-0.124\\
		$ $     &5.014&0.19,-0.33,-0.48,-0.01,-0.58,-0.16,0.06,-0.52 &16.086 &-0.269 &-0.134\\
		\bottomrule[0.5pt]\bottomrule[1.5pt]
	\end{tabular}
\end{table*}

\renewcommand{\tabcolsep}{0.4cm}
\renewcommand{\arraystretch}{0.9}
\begin{table*}[!htbp]
\caption{The final states with a decay width ratio of 1 are assumed to have the largest $|c_i|^2 \cdot k$ values, with the corresponding final states having $|c_i|^2 \cdot k$ values of approximately 0.1 GeV denoted in parentheses. A “0” indicates a branching ratio less than $1 \times 10^{-4}$, the star denotes states below the threshold, and “-” represents scattering states.}
	\label{table:decayssbbb}	
	\begin{tabular}{cc|cccc|cccc}
		\bottomrule[1.5pt]\bottomrule[0.5pt]
		&$ $ &$bbs\otimes s\bar{b}$ &$ $ &$ $ &$$&$css\otimes b\bar{b}$&$ $& & \\
		$J^{P}$ &Mass &$\Omega_{bb}^{\ast}\bar{B_{s}^{\ast}}$ &$\Omega_{bb}\bar{B_{s}^{\ast}}$ &$\Omega_{bb}^{\ast}\bar{B_{s}}$ &$\Omega_{bb}\bar{B_{s}}$ &$\Omega_{b}^{\ast}\Upsilon$
		&$\Omega_{b}^{\ast}\eta_{b}$ &$\Omega_{b}\Upsilon$&$\Omega_{b}\eta_{b}$\\ \hline
		${{5/2}^{-}}$
		&16.001  &\text{-}&$ $&$ $&$ $&$ $&$ $&$ $&$ $\\
		${{3/2}^{-}}$
		&16.229  &1&0.572&0.393&&0.569&&1&0.387\\ 
		&16.124  &0.116&1&0.074&&1&&0.119&0.075\\ 
		&16.109  &1&0.206&0.069&&0.190&&0.586&1\\
		&16.095  &0.035&0.779&1&&0.786&&0.036&1\\
		${{1/2}^{-}}$
		&16.235&1&0.089&&0.180&0.233&0.073&1&\\ 
		&16.126&0.024&1&&0.269&0.095&0.017&1&\\ 
		&16.104&0.041&1&&0.149&0.351&1&0.107&\\
		&16.086&0.081&0.011&&1&1&0.736&0.003&\\
		\bottomrule[0.5pt]\bottomrule[1.5pt]
	\end{tabular}
\end{table*}

The numerical results for the $sscc\bar{c}$ and $ssbb\bar{b}$ systems with strangeness $S = -2$ are presented in Tables \ref{table:massssccc} and \ref{table:massssbbb}. For the $sscc\bar{c}$ system, the mass ranges from $6.18 \, \text{GeV}$ to $5.98 \, \text{GeV}$, with the states $\left( 3/2^{-}, 6.014 \right)$ and $\left( 1/2^{-}, 5.975 \right)$ showing significant binding energies and enhanced compact stability. In contrast, the mass distribution for the $ssbb\bar{b}$
system spans from $16.24\, \text{GeV}$ to $16.00\, \text{GeV}$, with binding energies typically around $-120 \, \text{MeV}$. Additionally, the decay width information for both the $sscc\bar{c}$
and $ssbb\bar{b}$ systems is shown in Tables \ref{table:decayssccc} and \ref{table:decayssbbb}.

\subsection{$qQQQ\bar{Q}$ System}
\label{sec:qQQQQ}

In this section, we present information on the bag radius, mass spectrum, magnetic moment, and bag binding energy of the $nQQQ\bar{Q}$ system, specifically including $nccc\bar{c}$ and $nbbb\bar{b}$, as well as the strangeness $S = -1$ systems $sccc\bar{c}$ and $sbbb\bar{b}$. For the $nccc\bar{c}$
system, the bag radius is approximately $5.2 \, \text{GeV}$, with the mass distribution ranging from $6.93 \, \text{GeV}$ to $7.12 \, \text{GeV}$. The bag binding energy is around $-100 \, \text{MeV}$, indicating a deep binding. Additionally, the calculated mass of $nccc\bar{c}$ exceeds the thresholds of $\Xi_{cc}^{\ast} J/\psi$ and $\Omega_{ccc} D^{\ast}$, where the masses of $\Xi_{cc}^{\ast}$ and $\Omega_{ccc}$ are 3.714 GeV and 4.841 GeV, respectively \cite{Zhang:2021yul}. For the $nbbb\bar{b}$ system, the bag radius is approximately $4.3 \, \text{GeV}$. Due to the color-magnetic interactions of the $b$ quark, the mass distribution is confined to a very narrow range between $20.36 \, \text{GeV}$ and $20.42 \, \text{GeV}$, with a bag binding energy of about $-270 \, \text{MeV}$. The masses are also above the thresholds of $\Xi_{bb}^{\ast} \Upsilon$ and $\Omega_{bbb} B^{\ast}$, where the masses of $\Xi_{bb}^{\ast}$ and $\Omega_{bbb}$ are 10.360 GeV and 14.626 GeV, respectively. For the $sccc\bar{c}$ system, the bag radius is comparable to that of $nccc\bar{c}$, and there is little sensitivity of the bag radius and binding energy to the presence of strange quarks. The mass distribution lies between $7.06 \, \text{GeV}$ and $7.22 \, \text{GeV}$. The mass range for $sbbb\bar{b}$ is between $20.44 \, \text{GeV}$ and $20.51 \, \text{GeV}$. Furthermore, the masses of both $sccc\bar{c}$ and $sbbb\bar{b}$ are above the thresholds of the corresponding mesons and baryons.

\renewcommand{\tabcolsep}{0.4cm}
\renewcommand{\arraystretch}{0.9}
\begin{table*}[!htbp]
	\caption{The bag radius $R$ of the pentaquark state $nccc\bar{c}$
		is measured in units of $\text{GeV}^{-1}$, with the representation of the eigenbasis given by $cc/bbn\otimes n\bar{c}/\bar{b}$, the mass spectrum is expressed in units of GeV, the magnetic moment is given in units of $\mu_N$, and the binding energy $E_B$ is measured in units of GeV.}
	\begin{tabular}{ccccccc}
		\bottomrule[1.5pt]\bottomrule[0.5pt]
		State&$J^{P}$ &$R_{0}$&Eigenvector($cc/bbn\otimes c\bar{c}/b\bar{b}$) &$M_{bag}$ &$\mu_{bag}$&$E_{B}$\\ \hline
		$nccc\bar{c}$&${5/2}^{-}$     &5.298&-   &7.065 &2.891,0.056&-0.072\\
		&${3/2}^{-}$     &5.315&0.44,-0.14,-0.31,-0.67,-0.30,0.38,0.08&7.085 &1.311,0.047&-0.068\\
		&$ $     &5.273&-0.51,-0.13,-0.61,-0.25,0.45,-0.09,0.27&7.039 &2.380,-0.281 &-0.077\\
		&$ $     &5.084&-0.21,0.62,0.38,-0.53,-0.05,-0.26,0.26&6.927 &1.346,1.306 &-0.119\\
		&${1/2}^{-}$     &5.398&0.29,-0.44,-0.04,-0.32,0.57,0.54,0.04,-0.03 &7.122 &1.105,0.556 &-0.050\\
		&$ $     &5.246&0.07,-0.49,0.37,0.24,-0.54,0.33,-0.34,0.23&7.023 &1.165,0.302&-0.083\\
		&$ $     &5.158&0.64,0.25,0.10,0.29,-0.04,0.07,0.51,0.41&6.975 &-0.630,-0.166&-0.103\\
		$nbbb\bar{b}$&${5/2}^{-}$     &4.385&-   &20.401 &1.383,-0.948&-0.263\\
		&${3/2}^{-}$     &4.378&-0.45,0.15,0.31,0.65,0.30,-0.39,-0.07&20.412 &0.677,-0.458&-0.265\\
		&$ $     &4.324&0.51,0.12,0.61,0.27,-0.45,0.10,-0.27&20.386 &1.337,-0.819 &-0.276\\
		&$ $     &4.205&-0.19,0.62,0.38,-0.54,-0.07,-0.25,0.26&20.341&-0.253,-0.218 &-0.300\\
		&${1/2}^{-}$     &4.411&0.28,-0.43,-0.06,-0.33,0.59,0.53,0.05,-0.04 &20.422&0.107,-0.248 &-0.258 \\
		&$ $     &4.307&0.21,-0.45,0.38,0.28,-0.51,0.36,-0.22,0.31&20.381 &0.179,-0.277&-0.279\\
		&$ $     &4.259&0.62,0.35,0.02,0.23,0.07,0.01,0.56,-0.35&20.362&0.058,0.096&-0.289\\
		\bottomrule[0.5pt]\bottomrule[1.5pt]
	\end{tabular}
\end{table*}

\renewcommand{\tabcolsep}{0.45cm}
\renewcommand{\arraystretch}{0.9}
\begin{table*}[!htbp]
	\caption{The bag radius $R$ of the pentaquark state $sccc\bar{c}$
		is measured in units of $\text{GeV}^{-1}$, with the representation of the eigenbasis given by $cc/bbs\otimes s\bar{c}/\bar{b}$, the mass spectrum is expressed in units of GeV, the magnetic moment is given in units of $\mu_N$, and the binding energy $E_B$ is measured in units of GeV.}
	\begin{tabular}{ccccccc}
		\bottomrule[1.5pt]\bottomrule[0.5pt]
		State&$J^{P}$ &$R_{0}$&Eigenvector($cc/bbs\otimes c\bar{c}/b\bar{b}$) &$M_{bag}$ &$\mu_{bag}$&$E_{B}$\\ \hline
		$sccc\bar{c}$&${5/2}^{-}$     &5.324&-   &7.174 &0.282& -0.066 \\
		&${3/2}^{-}$     &5.339&-0.044,0.13,0.30,0.68,0.30,-0.38,-0.08&7.188& 0.161&-0.063 \\
		&$ $     &5.303&0.51,0.14,0.61,0.25,-0.45,0.09,-0.27&7.155 & -0.067&-0.071\\
		&$ $     &5.127&-0.22,0.62,0.38,-0.53,-0.05,-0.27,0.26&7.059&1.295&-0.109\\
		&${1/2}^{-}$     &5.419&0.29,-0.45,-0.04,-0.31,0.56,0.55,0.04,-0.02 &7.224 &0.582 &-0.045\\
		&$ $     &5.279&0.02,-0.50,0.36,0.22,-0.54,0.32,-0.37,0.20&7.141 &0.289&-0.076\\
		&$ $     &5.192&0.65,0.22,0.13,0.30,-0.08,0.09,0.48,0.42&7.098 &-0.105&-0.095\\
		$sbbb\bar{b}$&${5/2}^{-}$     &4.419&-   &20.491 &-0.798&-0.257\\
		&${3/2}^{-}$     &4.389&-0.45,0.15,0.31,0.66,0.30,-0.39,-0.07&20.504 &-0.382&-0.263\\
		&$ $     &4.392&0.51,0.12,0.61,0.27,-0.45,0.10,-0.27&20.482 &-0.684 &-0.262\\
		&$ $     &4.286&-0.19,0.62,0.38,-0.54,-0.07,-0.25,0.26&20.442&-0.222&-0.284\\
		&${1/2}^{-}$     &4.470&0.28,-0.43,-0.06,-0.33,0.59,0.43,0.05,-0.04&20.513 &-0.227 &-0.246\\
		&$ $     &4.377&0.19,-0.45,0.38,0.27,-0.51,0.36,-0.24,0.30&20.478 &-0.256&-0.265\\
		&$ $     &4.334&0.62,0.33,0.03,0.24,0.05,0.01,0.56,0.36&20.461 &0.103&-0.271\\
		\bottomrule[0.5pt]\bottomrule[1.5pt]
	\end{tabular}
\end{table*}

\begin{figure}[t]
	\centering
	\includegraphics[width=0.45\textwidth]{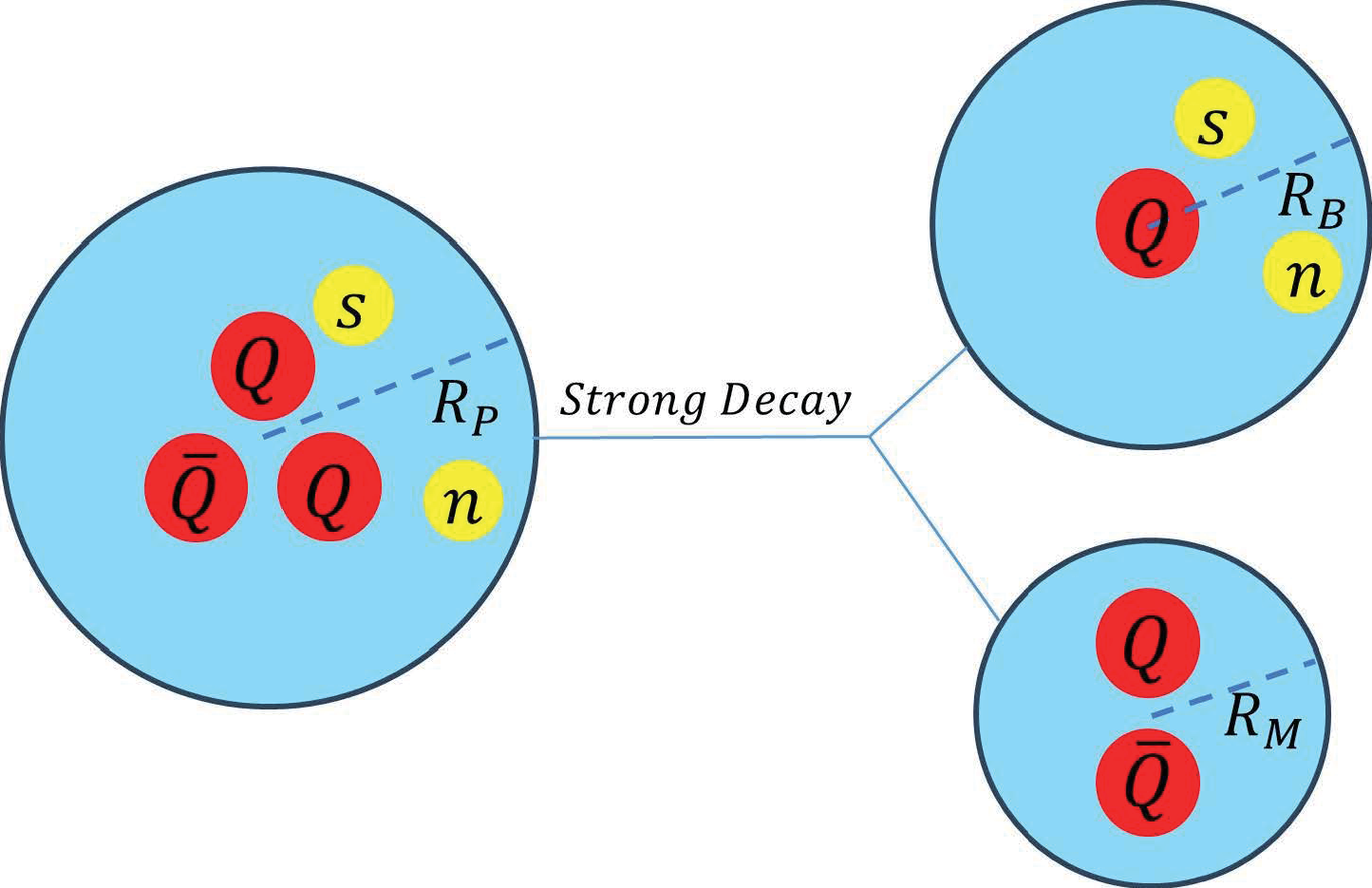}
	\caption{Distribution of heavy quarks and decay schematic in triply heavy pentaquark states, where $R_{i}$ (with $i = P, B, M$) represents the radius of the pentaquark state, baryon radius, and meson radius, respectively.}
	\label{fig:qqQQQ}
\end{figure}

Based on the results from the calculations in Sects. \ref{sec:qqQQQ} and \ref{sec:qQQQQ}, we will provide a qualitative discussion on multi-flavor pentaquark states, focusing on two themes:

I. The compactness potential of multi-flavor pentaquark states.

II. The threshold characteristics and strong decay properties of multi-flavor pentaquark states.

As shown in Table \ref{tab:radii}, the impact of light-flavor quarks on the hadronic bag radius is significantly larger than that of heavy-flavor quarks. On the other hand, considering the short-range interactions between heavy quarks, these interactions can affect the internal dynamics of hadrons in multi-heavy flavor systems. Therefore, in the case of triply heavy pentaquarks, the short-range interactions between heavy quarks may lead to the formation of a core (see Fig. \ref{fig:qqQQQ}). In addition, the intentional for compactness in multi-flavor pentaquark states is supported by the bag binding energy.

In our calculations, nearly all multi-flavor pentaquark states have at least one two-body strong decay channel. We provide some qualitative explanations based on the bag model to explain why compact structures exhibit threshold characteristics. For pentaquark systems that satisfy the binding energy condition $E_{B} < 0$
, although they may possess the intentional for compactness, their compact stability is lower than that of traditional mesons or baryons. As shown in Fig. \ref{fig:qqQQQ}, the average radius $R_{P} = 5.516 \, \text{GeV}^{-1}$ for the pentaquark state $nscc\bar{c}$ (ignoring chromomagnetic interactions), and according to Eq. (\ref{EB}), its binding energy is $E_{B} = -23 \, \text{MeV}$. For the baryon $nsc$, the average radius is $R_{B} = 4.86 \, \text{GeV}^{-1}$, with a binding energy of $E_{B} = -166 \, \text{MeV}$. For the meson $c\bar{c}$, the average radius is $R_{M} = 3.4 \, \text{GeV}^{-1}$, with a binding energy of $E_{B} = -471 \, \text{MeV}$. Therefore, the strong decay processes of compact pentaquark states can be approximately understood in the bag model framework as a process of releasing binding energy, leading to a direction toward higher compact stability.

\section{Summary}
\label{sec:summary}

In this study, we investigated the limiting binding radius under color interactions within the framework of the MIT bag model, which is set at $5.615 \, \text{GeV}^{-1}$. We extended this result to multi-quark systems to evaluate their potential for forming compact structures. In our study of pentaquarks, we examined several combinations of heavy and light flavors and found that the flavor combination $nnnc\bar{c}$
has a bag radius that is close to and slightly above the binding limit, while the average radius for the flavor combination $nncc\bar{c}$
is below the binding limit. This suggests that there may exist pentaquark states with higher compact stability in the $nncc\bar{c}$ system. Subsequently, we conducted a systematic study of pentaquarks containing three or four heavy quarks using the bag model. We took chromomagnetic interactions into account and calculated the radius, magnetic moment, mass, and bag binding energy for both the triply and quadruple-heavy flavor pentaquarks. Additionally, we examined the partial width ratios of the triply-heavy flavor pentaquarks in the baryon-meson representation.

For the $nncc\bar{c}$ system, the predicted mass range is $5.53-6.00 \, \text{GeV}$. After considering the chromomagnetic interactions, the binding energies of most states are above $-20 \, \text{MeV}$. Additionally, the mass distribution ensures that each state has at least one strong decay channel, although for some decay channels, the decay width is suppressed due to phase space or final state occupancy effects. For the $nnbb\bar{b}$ system, the mass range is distributed between $15.60\, \text{GeV}$ and $16.10 \, \text{GeV}$. All binding energies are above $-120 \, \text{MeV}$, indicating a strong binding. Furthermore, we have calculated the triplet or quadruplet heavy pentquarks with strangeness quantum numbers $S = -1, -2$, and we have also predicted the magnetic moments for each state. Our theoretical rationale for the potential compact configurations of multiquark states, particularly pentaquarks, is based on two main aspects: first, the bag binding energy defined by the bag model indicates the presence of compact stability; second, we take into account the short-range interactions between heavy quarks.

One insight from the bag model is that compact hadronic states have a limiting radius, which could be related to the confinement properties of hadrons. The study of multiquark states can provide evidence for this conclusion, particularly since triplet pentaquarks may exist at a relatively sensitive scale. Additionally, we look forward to more extensive studies of multi-quark states from lattice QCD, as well as richer experimental data to support our conclusions.

\medskip
\textbf{ACKNOWLEDGMENTS}

We thanks Kai-Kai Zhang and Ming Zhu Liu for usefull discussings. 
The work is supported by the National Natural Science Foundation of China (Grants No. 12475026 and No.12075193).

\appendix
\section{Color and Spin Wavefunctions}
\label{apd:WF}

For the pentaquark state, the direct product among the five valence quarks can be considered as the product of a group formed by three valence quarks and a quark-antiquark pair, leading to the following expression:
\begin{equation}
(3\otimes3\otimes3)\otimes(3\otimes\bar{3})=(1\oplus8\oplus8\oplus10)\otimes(1\oplus8).
\end{equation}
On the right side of the above equation, there are two color combinations of $8 \otimes 8$, which can each yield a color singlet. Additionally, $1 \otimes 1$ can also produce a color singlet combination. Therefore, the pentaquark state can be characterized by a total of three types of color singlet wave functions, as shown in the following expression:

\begin{equation}\label{8b1}
    \begin{aligned}
        \phi_{1}^{P} &=\frac{1}{4\sqrt{3}}\Big[(2bbgr-2bbrg+gbrb-gbbr+bgrb-bgbr\\
        &-rbgb+rbbg-brgb+brbg)\bar{b}+(2rrbg-2rrgb\\
        &+rgrb-rgbr+grrb-grbr+rbgr-rbrg+brgr\\
        &-brrg)\bar{r}+(2ggrb-2ggbr-rggb+rgbg-grgb\\
        &+grbg+gbgr-gbrg+bggr-bgrg)\bar{g}\Big],
    \end{aligned}
\end{equation}
    
\begin{equation}\label{8b2}
    \begin{aligned}
        \phi_{2}^{P} &=\frac{1}{12}\Big[(3bgbr-3gbbr-3brbg+3rbbg-rbgb-2rgbb \\
        &+2grbb+brgb+gbrb-bgrb)\bar{b}+(3grrb-3rgrb \\
        &-3brrg+3rbrg-rbgr-2gbrr+2bgrr-grbr  \\
        &+rgbr+brgr)\bar{r}+(3grgb-3rggb+3bggr-3gbgr \\
        &-grbg+rgbg+2rbgg-2brgg+gbrg-bgrg)\bar{g}\Big],    
    \end{aligned}
\end{equation}
    
\begin{equation}\label{8b3}
    \begin{aligned}
        \phi_{3}^{P} &=\frac{1}{3\sqrt{2}}\Big[(grbb-rgbb+rbgb-brgb+bgrb-gbrb)\bar{b}\\
        &+(grbr-rgbr+rbgr-brgr+bgrr-gbrr)\bar{r} \\
        &+(grbg-rgbg+rbgg-brgg+bgrg-gbrg)\bar{g}\Big].    
    \end{aligned}
\end{equation}

For the spin basis of the pentaquarks, a similar treatment can be applied, which can be regarded as the spin summation of the three-quark group and the quark-antiquark pair:

\begin{equation}\label{Spin}
	\begin{aligned}
		\chi_{1}^{P}&=|[(12)_{1}3]_{3/2}(4\bar{5})_{1}\rangle_{5/2}, 
		\chi_{2}^{P}=|[(12)_{1}3]_{3/2}(4\bar{5})_{1} \rangle_{3/2},\\
		\chi_{3}^{P}&=|[(12)_{1}3]_{3/2}(4\bar{5})_{0}\rangle_{3/2}, 
		\chi_{4}^{P}=|[(12)_{1}3]_{1/2}(4\bar{5})_{1} \rangle_{3/2},\\
		\chi_{5}^{P}&=|[(12)_{0}3]_{1/2} (4\bar{5})_{1}\rangle_{3/2}, 
		\chi_{6}^{P}=|[(12)_{1}3]_{3/2} (4\bar{5})_{1}\rangle_{1/2},\\
		\chi_{7}^{P}&=|[(12)_{1}3]_{1/2} (4\bar{5})_{1}\rangle_{1/2},  
		\chi_{8}^{P}=|[(12)_{1}3]_{1/2}(4\bar{5})_{0}\rangle_{1/2},\\
		\chi_{9}^{P}&=|[(12)_{0}3]_{1/2}(4\bar{5})_{1}\rangle_{1/2}, 
		\chi_{10}^{P}=|[(12)_{0}3]_{1/2} (4\bar{5})_{0}\rangle_{1/2}.
	\end{aligned}
\end{equation}
Here, the indices 1, 2, 3, 4, and 5 refer to the quarks, while the subscripts in parentheses indicate the spins of the respective subgroups. The outermost subscript denotes the total spin of the five-quark state, which is crucial for determining the decay final states discussed in the text.

The basis vectors of the five-quark states are described by the algebra of the SU(3) $\otimes$ SU(2) groups, resulting in a total of thirty basis vectors, which are labeled according to their angular momentum. The basis vectors for angular momentum $J = \frac{5}{2}$, $J = \frac{3}{2}$, and $J = \frac{1}{2}$ are shown as follows.

\renewcommand{\tabcolsep}{0.1cm} \renewcommand{\arraystretch}{0.9}
\begin{table*}[!htb]
	\caption{Considering the flavor configuration, the color-spin eigenbasis of the pentaquark state}
	\label{tab:basis}
	\begin{tabular}{lccc}
		\bottomrule[1.5pt]\bottomrule[0.3pt]
		System &$I$ &$J^{P}$ &Color-spin wave functions \\ \hline
		$cnn\otimes c\bar{c}$/$bnn\otimes b\bar{b}$ & 0
		& ${5/2}^{-}$ & $\phi_{1}\chi_{1},\phi_{3}\chi_{1}$ \\
		&   & ${3/2}^{-}$ & $\phi_{1}\chi_{2},\phi_{1}\chi_{3},\phi_{1}\chi_{4},\phi_{2}\chi_{5},\phi_{3}\chi_{5}$ \\
		&   & ${1/2}^{-}$ & $\phi_{1}\chi_{6},\phi_{1}\chi_{7},\phi_{1}\chi_{8},\phi_{2}\chi_{9},\phi_{2}\chi_{10},\phi_{3}\chi_{9},\phi_{3}\chi_{10}$ \\
		 & 1
		& ${5/2}^{-}$ & $\phi_{1}\chi_{1},\phi_{3}\chi_{1}$ \\
		&   & ${3/2}^{-}$ & $\phi_{1}\chi_{5},\phi_{2}\chi_{2},\phi_{2}\chi_{3},\phi_{2}\chi_{4},\phi_{3}\chi_{2},\phi_{3}\chi_{3},\phi_{3}\chi_{4}$\\
		&   & ${1/2}^{-}$ & $\phi_{1}\chi_{9},\phi_{1}\chi_{10},\phi_{2}\chi_{6},\phi_{2}\chi_{7},\phi_{2}\chi_{8},\phi_{3}\chi_{6},\phi_{3}\chi_{7},\phi_{3}\chi_{8}$\\
		$ccn\otimes n\bar{c}$/$bbn\otimes n\bar{b}$&
		& ${5/2}^{-}$ & $\phi_{2}\chi_{1},\phi_{3}\chi_{1}$ \\
		&   & ${3/2}^{-}$ & $\phi_{1}\chi_{5},\phi_{2}\chi_{2},\phi_{2}\chi_{3},\phi_{2}\chi_{4},\phi_{3}\chi_{2},\phi_{3}\chi_{3},\phi_{3}\chi_{4}$ \\
		&   & ${1/2}^{-}$ & $\phi_{1}\chi_{9},\phi_{1}\chi_{10},\phi_{2}\chi_{6},\phi_{2}\chi_{7},\phi_{2}\chi_{8},\phi_{3}\chi_{6},\phi_{3}\chi_{7},\phi_{3}\chi_{8}$ \\
		
		$csn\otimes c\bar{c}$/$bsn\otimes b\bar{b}$&
		& ${5/2}^{-}$ & $\phi_{1}\chi_{1},\phi_{2}\chi_{1},\phi_{3}\chi_{1}$ \\
		&   & ${3/2}^{-}$ & $\phi_{1}\chi_{2},\phi_{1}\chi_{3},\phi_{1}\chi_{4},\phi_{1}\chi_{5},\phi_{2}\chi_{2},\phi_{2}\chi_{3},\phi_{2}\chi_{4},\phi_{2}\chi_{5},\phi_{3}\chi_{2},\phi_{3}\chi_{3},\phi_{3}\chi_{4},\phi_{3}\chi_{5}$ \\
		&   & ${1/2}^{-}$ & $\phi_{1}\chi_{6},\phi_{1}\chi_{7},\phi_{1}\chi_{8},\phi_{1}\chi_{9},\phi_{1}\chi_{10},\phi_{2}\chi_{6},\phi_{2}\chi_{7},\phi_{2}\chi_{8},\phi_{2}\chi_{9},\phi_{2}\chi_{10},\phi_{3}\chi_{6},\phi_{3}\chi_{7},\phi_{3}\chi_{8},\phi_{3}\chi_{9},\phi_{3}\chi_{10}$ \\
		
		$ccs\otimes n\bar{c}$/$ccn\otimes s\bar{c}$/$bbs\otimes n\bar{b}$/$bbn\otimes s\bar{b}$&
		& ${5/2}^{-}$ & $\phi_{2}\chi_{1},\phi_{3}\chi_{1}$ \\
		&   & ${3/2}^{-}$ & $\phi_{1}\chi_{5},\phi_{2}\chi_{2},\phi_{2}\chi_{3},\phi_{2}\chi_{4},\phi_{3}\chi_{2},\phi_{3}\chi_{3},\phi_{3}\chi_{4}$ \\
		&   & ${1/2}^{-}$ & $\phi_{1}\chi_{9},\phi_{1}\chi_{10},\phi_{2}\chi_{6},\phi_{2}\chi_{7},\phi_{2}\chi_{8},\phi_{3}\chi_{6},\phi_{3}\chi_{7},\phi_{3}\chi_{8}$ \\
		
		$css\otimes c\bar{c}$/$bss\otimes b\bar{b}$ &
		& ${5/2}^{-}$ & $\phi_{1}\chi_{1},\phi_{2}\chi_{1},\phi_{3}\chi_{1}$ \\
		&   & ${3/2}^{-}$ & $\phi_{1}\chi_{5},\phi_{2}\chi_{2},\phi_{2}\chi_{3},\phi_{2}\chi_{4},\phi_{3}\chi_{2},\phi_{3}\chi_{3},\phi_{3}\chi_{4}$\\
		&   & ${1/2}^{-}$ & $\phi_{1}\chi_{9},\phi_{1}\chi_{10},\phi_{2}\chi_{6},\phi_{2}\chi_{7},\phi_{2}\chi_{8},\phi_{3}\chi_{6},\phi_{3}\chi_{7},\phi_{3}\chi_{8}$\\
			
		$ccs\otimes s\bar{c}$/$bbs\otimes s\bar{b}$ &
		& ${5/2}^{-}$ & $\phi_{1}\chi_{1},\phi_{2}\chi_{1},\phi_{3}\chi_{1}$ \\
		&   & ${3/2}^{-}$ & $\phi_{2}\chi_{2},\phi_{2}\chi_{3},\phi_{2}\chi_{4},\phi_{3}\chi_{2},\phi_{3}\chi_{3},\phi_{3}\chi_{4}$\\
		&   & ${1/2}^{-}$ & $\phi_{1}\chi_{9},\phi_{1}\chi_{10},\phi_{2}\chi_{6},\phi_{2}\chi_{7},\phi_{2}\chi_{8},\phi_{3}\chi_{6},\phi_{3}\chi_{7},\phi_{3}\chi_{8}$\\	
			
		$cccn\bar{c}$/$bbbn\bar{b}$/$cccs\bar{c}$/$bbbs\bar{b}$ & 
		& ${5/2}^{-}$ & $\phi_{2}\chi_{1},\phi_{3}\chi_{1}$ \\
		&   & ${3/2}^{-}$ & $\phi_{1}\chi_{5},\phi_{2}\chi_{2},\phi_{2}\chi_{3},\phi_{2}\chi_{4},\phi_{3}\chi_{2},\phi_{3}\chi_{3},\phi_{3}\chi_{4}$ \\
		&   & ${1/2}^{-}$ & $\phi_{1}\chi_{9},\phi_{1}\chi_{0},\phi_{2}\chi_{6},\phi_{2}\chi_{7},\phi_{2}\chi_{8},\phi_{3}\chi_{6},\phi_{3}\chi_{7},\phi_{3}\chi_{8}$ \\
		\bottomrule[0.5pt]\bottomrule[1.5pt]
	\end{tabular}%
\end{table*}
$J^{P}=5/2^{-}$
\begin{equation}\label{cs1}
	\begin{aligned}
		\phi_{1}\chi_{1}=|[(12)^{6}_{1}3]^{8}_{3/2} (4\bar{5})^{8}_{1}\rangle_{5/2},\\
		\phi_{2}\chi_{1}=|[(12)^{\bar{3}}_{1}3]^{8}_{3/2} (4\bar{5})^{8}_{1}\rangle_{5/2},\\
		\phi_{3}\chi_{1}=|[(12)^{\bar{3}}_{1}3]^{1}_{3/2} (4\bar{5})^{1}_{1}\rangle_{5/2}.
	\end{aligned}
\end{equation}

$J^{P}=3/2^{-}$

\begin{equation}\label{cs2}
	\begin{aligned}
		\phi_{1}\chi_{2}=|[(12)^{6}_{1}3]^{8}_{3/2} (4\bar{5})^{8}_{1}\rangle_{3/2},\\
		\phi_{1}\chi_{3}=|[(12)^{6}_{1}3]^{8}_{3/2} (4\bar{5})^{8}_{0}\rangle_{3/2},\\
		\phi_{1}\chi_{4}=|[(12)^{6}_{1}3]^{8}_{1/2} (4\bar{5})^{8}_{1}\rangle_{3/2},\\
		\phi_{1}\chi_{5}=|[(12)^{6}_{0}3]^{8}_{1/2} (4\bar{5})^{8}_{1}\rangle_{3/2},\\
		\phi_{2}\chi_{2}=|[(12)^{\bar{3}}_{1}3]^{8}_{3/2} (4\bar{5})^{8}_{1}\rangle_{3/2},\\
		\phi_{2}\chi_{3}=|[(12)^{\bar{3}}_{1}3]^{8}_{3/2} (4\bar{5})^{8}_{0}\rangle_{3/2},\\
		\phi_{2}\chi_{4}=|[(12)^{\bar{3}}_{1}3]^{8}_{1/2} (4\bar{5})^{8}_{1}\rangle_{3/2},\\
		\phi_{2}\chi_{5} = | [(12)^{\bar{3}}_{0}3]^{8}_{1/2} (4\bar{5})^{8}_{1} \rangle_{3/2},\\
		\phi_{3}\chi_{2} = | [(12)^{\bar{3}}_{1}3]^{1}_{3/2} (4\bar{5})^{1}_{1} \rangle_{3/2},  \\
		\phi_{3}\chi_{3} = | [(12)^{\bar{3}}_{1}3]^{1}_{3/2} (4\bar{5})^{1}_{0} \rangle_{3/2},  \\
		\phi_{3}\chi_{4} = | [(12)^{\bar{3}}_{1}3]^{1}_{1/2} (4\bar{5})^{1}_{1} \rangle_{3/2},  \\
		\phi_{3}\chi_{5} = | [(12)^{\bar{3}}_{0}3]^{1}_{1/2} (4\bar{5})^{1}_{1} \rangle_{3/2}.
	\end{aligned}
\end{equation}

$J^{P}=1/2^{-}$

\begin{equation}\label{cs3}
	\begin{aligned}
		&\phi_{1}\chi_{6} = | [(12)^{6}_{1}3]^{8}_{3/2} (4\bar{5})^{8}_{1} \rangle_{1/2},  \\
		&\phi_{1}\chi_{7} = | [(12)^{6}_{1}3]^{8}_{1/2} (4\bar{5})^{8}_{1} \rangle_{1/2},  \\
		&\phi_{1}\chi_{8} = | [(12)^{6}_{1}3]^{8}_{1/2} (4\bar{5})^{8}_{0} \rangle_{1/2},  \\
		&\phi_{1}\chi_{9} = | [(12)^{6}_{0}3]^{8}_{1/2} (4\bar{5})^{8}_{0} \rangle_{1/2},  \\
		&\phi_{1}\chi_{10} = | [(12)^{6}_{0}3]^{8}_{1/2} (4\bar{5})^{8}_{0} \rangle_{1/2},  \\    
		&\phi_{2}\chi_{6}  = | [(12)^{\bar{3}}_{1}3]^{8}_{3/2} (4\bar{5})^{8}_{1} \rangle_{1/2}, \\
		&\phi_{2}\chi_{7}  = | [(12)^{\bar{3}}_{1}3]^{8}_{1/2} (4\bar{5})^{8}_{1} \rangle_{1/2}, \\
		&\phi_{2}\chi_{8}  = | [(12)^{\bar{3}}_{1}3]^{8}_{1/2} (4\bar{5})^{8}_{0} \rangle_{1/2}, \\
		&\phi_{2}\chi_{9}  = | [(12)^{\bar{3}}_{0}3]^{8}_{1/2} (4\bar{5})^{8}_{1} \rangle_{1/2},  \\
		&\phi_{2}\chi_{10} = | [(12)^{\bar{3}}_{0}3]^{8}_{1/2} (4\bar{5})^{8}_{0} \rangle_{1/2},  \\
		&\phi_{3}\chi_{6}  = | [(12)^{\bar{3}}_{1}3]^{1}_{3/2} (4\bar{5})^{1}_{1} \rangle_{1/2}, \\
		&\phi_{3}\chi_{7}  = | [(12)^{\bar{3}}_{1}3]^{1}_{1/2} (4\bar{5})^{1}_{1} \rangle_{1/2}, \\
		&\phi_{3}\chi_{8} = | [(12)^{\bar{3}}_{1}3]^{1}_{1/2} (4\bar{5})^{1}_{0} \rangle_{1/2},\\
		&\phi_{3}\chi_{9}  = | [(12)^{\bar{3}}_{0}3]^{1}_{1/2} (4\bar{5})^{1}_{1} \rangle_{1/2},  \\
		&\phi_{3}\chi_{10} = | [(12)^{\bar{3}}_{0}3]^{1}_{1/2} (4\bar{5})^{1}_{0} \rangle_{1/2}.
	\end{aligned}
\end{equation}

Finally, after considering flavor symmetry and the Pauli exclusion principle, the color-spin eigenbasis of the pentaquarks corresponding to the flavor combinations discussed in this investigation are presented in Table \ref{tab:basis}.

\end{document}